# Direct Air Capture in Europe's 2050 Energy System: Integration, Storage and Cost Drivers

Maximilian Bernecker[1]*, Felix Müsgens[1]

[1]Chair of Energy Economics, Brandenburg University of Technology Cottbus–Senftenberg (BTU), Cottbus, Brandenburg, Germany

Correspondence: maximilian.bernecker@b-tu.de

**Abstract:** Direct Air Carbon Capture and Storage (DACCS) can balance residual hard-to-abate greenhouse gas emissions, for example from air transport, the cement industry, and livestock agriculture. Today, there is a wide variety of cost estimates for DACCS, driven, to a significant extent, by differences in electricity cost. A holistic view on electricity costs requires the integration of DACCS into long-term energy system models. This paper distinguishes the different DACCS components (direct air capturing, carbon transport, and carbon storage) and integrates them into a European capacity expansion model for a fully decarbonised electricity system in 2050. Empirically, we explore how two dimensions affect the total system costs of DACCS. The first dimension varies the availability of $CO_2$ storage locations. One alternative restricts storage locations to offshore in the North Sea, specifically depleted natural gas fields. The other alternative allows storage in suitable locations across Europe, including onshore sites. We find that limiting $CO_2$ storage to North Sea sites increases overall capture costs by approximately 10 %. The second dimension varies DACCS integration into the electricity system. We differentiate between three alternatives: fully isolated, fully integrated, and retrospectively added to an existing system. We find that neglecting system integration – i.e. treating direct air capture system as a stand-alone technology – increases capture costs by up to 30 %.

# 1 INTRODUCTION

The Intergovernmental Panel on Climate Change (IPCC) stated that "carbon dioxide removal"[1] is necessary for achieving a net-zero emission goal (IPCC, 2023). The process of removing $CO_2$ directly from the air and storing it permanently is referred to as Direct Air Carbon Capture and Storage (DACCS, see Erans et al. (2022) for an overview of the technology). Recent literature highlights the importance of DACCS to meet climate targets (Hanna et al. 2021; Galán-Martín et al. 2021; Kazlou et al. 2024). To promote investments and support the rollout of carbon management technologies in Europe, the European Union (EU) published the industrial carbon management strategy in 2024, whose aim is to capture 450 Mt $CO_2$ per year by 2050, from which around 135 Mt (30%) is estimated to come from direct air capture. Other European countries, e.g. the United Kingdom and Norway, are also pursuing ambitious carbon removal strategies (Department for Energy Security & Net Zero, 2023, Carbon Removal in Norway – National Policy Overview, 2025). Hence, both politics and academia (Marcucci et al. 2017; Fasihi et al. 2019; Galán-Martín et al. 2021; IPCC 2023) agree that DACCS can contribute significantly to the decarbonisation of the economy, especially in hard-to-abate sectors.

To what extent this contribution manifests depends on the cost-competitiveness of DACCS, which drives both investor interest and policy support. Although considerable efforts have been made in recent years to estimate the cost for DACCS, studies' estimates vary at least by a factor of ten. For example, a review by Erans et al. (2022) reports cost estimates for 2050 ranging from 40 to 80/t $CO_2$ in optimistic scenarios, but point out that 400 to 800 EUR/t $CO_2$ is mentioned in other studies. While uncertainty around the long-term deployment of new technologies is inherently large, key drivers to the uncertainty of overall DACCS cost estimates can be identified and explored. Our work quantifies and explains the contribution of two key drivers– electricity prices and positioning of carbon storage sites.

We focus on electricity prices and the impact of positioning storage sites for several reasons. First, electricity prices drive the operational expenditures (OPEX) of direct air capture and are thus important for total DACCS costs. The literature estimates costs for electricity can account for nearly half of total DACCS costs (Fasihi et al. 2019; Lux et al. 2023; Sievert et al. 2024). Second, most existing literature either uses exogenous electricity prices, i.e. assumptions based on the literature, or determines the electricity cost in an isolated system wherein direct air capture is the only consumer. This is a simplifying assumption, as DACCS will be added to existing systems in most cases, with DACCS affecting the rest of the system and vice versa. Moreover, assumed prices vary widely, ranging from 5 EUR/ $MWh_{el}$ to 133 EUR/ $MWh_{el}$ (Young et al. 2023; Terlouw et al. 2024). Third, the controversial impact of storage sites is discussed, e.g. whether $CO_2$ storage is permitted only offshore or also onshore (IEA, 2020; Chan et al., 2024; Flöer et al., 2025). Furthermore, to the best of our knowledge, there is no study that quantifies the impact of different $CO_2$ storage locations on the underlying energy system and the DACCS cost.

[1] Carbon dioxide removal is also referred to as negative emissions, because it reduces the amount of $CO_2$ in the atmosphere.

We develop and apply a detailed model of a 2050 decarbonised European electricity system that optimises carbon capture, transport, and storage infrastructure. As low-temperature (LT) DACCS systems are frequently identified to be technologically and economically competitive (Fasihi et al. 2019; Sabatino et al. 2021; Young et al. 2023; Sievert et al. 2024; Terlouw et al. 2024), our analysis focuses on LT DACCS. We model investment in and dispatch of electricity generation and transmission technologies, deriving the DACCS's OPEX endogenously. We explore the impact of three different modelling approaches on the DACCS costs and analyse how they influence the cost-optimal energy system layout. First, we analyse the isolated electricity demand from DACCS. Second, we compute at the additional cost when DACCS are added to an existing, optimised electricity system. Third, we analyse a fully integrated system, where DACCS are part of the cost-optimal system composition from the beginning.

Therefore, we can compare six distinct scenarios: two siting alternatives multiplied by three system integration alternatives. Specifically, we:

- Compare isolated (stand-alone) direct air capture system configurations with integrated deployment strategies in the electricity system planning.
- Assess the impact of different $CO_2$ storage locations, contrasting North Sea-based storage options with onshore storage potentials across Europe.

Our results reveal infrastructure trade-offs between regional renewable energy use, electricity transmission, and DACCS system integration, allowing us to quantify the electricity system's contribution to negative emissions in terms of both total system cost and cost per ton of $CO_2$ removed. Explaining cost differences and reducing uncertainty helps in assessing the role of DACCS in the European energy system: variations of more than one order of magnitude are an obvious challenge for investors and policy advisors when deciding on investment and potential subsidies for technology pathways.

In section 2 we provide a literature overview on the need for direct air capture systems as negative emission technology, and on techno-economical studies to assess their future cost, highlighting the research gap. Section 3 presents our methodology by introducing the underlying energy system model, data, and scenario assumptions. The results are presented in section 4, followed by a discussion and conclusion in section 5.

# 2 LITERATURE

In the following literature review, subsection 2.1 summarises and synthesises existing estimates of future long-term LT DACCS costs. Subsections 2.2 and 2.3 examine two key drivers underlying this uncertainty: electricity prices and the role of $CO_2$ storage location. Finally, subsection 2.4 outlines the research gap and situates our contribution within the existing literature.

## 2.1 Cost Estimates of DACCS

The total capturing costs of LT DACCS systems can be separated into two components: the costs of the direct air capture units (the DAC part of DACCS) and the costs for $CO_2$ transport and storage (the CS part). As the DAC part has higher costs associated (>90 % of total costs, see e.g. Fasihi et al. 2019; Lux et al. 2023; Terlouw et al. 2024), we analyse the literature on that part, further differentiating between capital expenditures (CAPEX) and operational expenditures (OPEX) of the direct air capture units.

Studies estimating the future CAPEX of direct air capture units focus strongly on technological learning curves, equipment costs, and scale effects. Significant uncertainty lies in the unknown deployment scale (Young et al. 2023), making assumptions regarding the latter effect vital to the analysis. Depending on future deployment, it remains unclear whether the CAPEX will continue to be the dominating cost factor (Sabatino et al. 2021) or whether its relative contribution will fall below the OPEX over time (Lux et al. 2023). A review of 31 studies by Wenzel et al. (2025b) highlights this uncertainty by providing an overview of the expected future purchase equipment costs for LT DAC units. For 2050, the median estimate is 105 EUR/ $tCO_2$ with a substantial dispersion across studies, reflected in an interquartile range of 253 EUR/ $tCO_2$. A key in reducing the uncertainty lies in an acceleration of the DACCS capacity scale-up, as learning effects and technological maturity are expected to reduce the capital cost (Young et al. 2023). Overall, depending on the location and energy supply source, the share of DAC CAPEX from the total DACCS cost are estimated in the literature to be between 10 % and 54 %, (Lux et al. 2023; Sievert et al. 2024; Terlouw et al. 2024).

Another major component of capture costs is the operating expenditures (OPEX) of direct air capture units. OPEX consist of three components: fixed operation and maintenance costs, cost for heat and cost for electricity. Fixed operation and maintenance costs are typically assumed to be comparably low and often directly linked to the DAC CAPEX. For instance, Fasihi et al. (2019) and Lux et al. (2023) assume that annual operation and maintenance costs amount to 4 % of the DAC CAPEX. The cost for heat depends on the assumptions and the context of the paper. When a point source for heat is assumed to be available, the cost is assumed to be negligible, as Fasihi et al., (2019) and Terlouw et al. (2024), for example, assume the use of waste heat at zero cost. Other studies assume the heat energy must be generated within the DACCS process. In that case, the energy requirements of low-temperature DACCS mainly comprise electricity demand for the DAC units' air contactors, vacuum pumps, and compressors; as well as low-temperature heat demand for sorbent regeneration. This heat is typically supplied via heat pumps, which provide the thermal energy via electricity consumption.

In that set-up, the DAC OPEX are essentially determined by the cost for electricity, both directly (e.g. for ventilation) and indirectly (for heating the sorbent via heat pumps). Hence, an analysis of electricity price assumptions is vital for a cost assessment of DACCS. Note that the costs for electricity are the product of an electricity price with the amount of electricity consumed in the DACCS process. The literature tends to differ more in the electricity price assumptions than in the amount of electricity consumed (Hanna et al. 2021; Ozkan et al. 2022). The OPEX estimates for DACCS systems in the year 2050 vary widely in the literature. Fasihi et al. (2019) estimate that OPEX account for around 60 % of total DACCS costs, driven primarily by the operation of the DAC units, with electricity consumption alone representing 48 %. Lux et al. (2023) report an even higher contribution, with DAC-related OPEX constituting roughly 75-82 % of total DACCS costs, of which 64-74 % are attributed to electricity expenditures. Sievert et al. (2024) estimate a lower OPEX share of 36-40 %, though still mainly determined by the electricity consumption of the DAC unit.

Finally, the $CO_2$ transport and storage infrastructure also contributes to the overall DACCS cost but is found to be a less-dominant cost factor, with its share of the overall DACCS cost often estimated below 10 % (Fasihi et al. 2019; Lux et al. 2023; Terlouw et al. 2024). Nonetheless, the representation of the $CO_2$ infrastructure influences the other components of the DACCS supply, in particular the siting of direct air capture units and their associated energy supply. In the existing literature, most studies focus on other aspects and treat the $CO_2$ infrastructure in a highly simplified manner, e.g. omitting the impact of regulatory regimes and instead applying uniform, generic cost assumptions per ton of $CO_2$ stored or transported (Rubin et al. 2015; Breyer et al. 2020; Smith et al. 2021; Lux et al. 2023). This simplification does not take into account the geospatial scale that shapes both electricity system operation and $CO_2$ transport and storage networks.

As our analysis focuses on the cost for electricity consumption of DACCS and the representation of the $CO_2$ sector, we discuss both in more detail in the following two subsections.

## 2.2 Electricity Costs of DACCS

As outlined in the previous section, cost for electricity is a vital parameter for the total cost assessment of DACCS. When heat for LT DAC is generated with heat pumps, the vast majority of OPEX are driven by the cost of electricity, making up between 36-74 % of the total DACCS cost. Furthermore, we find a huge variance in the assumed electricity cost between studies. We have already pointed out that this uncertainty is mostly driven by the assumed price per unit of electricity and less by the amount of electricity consumed in the DAC process (Hanna et al., 2021; Ozkan et al., 2022; Young et al., 2023; Wenzel et al. 2025b).

The huge differences in assumed electricity prices result partly from fundamental differences in methodology. Methodologies can be broadly categorised into three approaches: (i) the use of exogenous electricity price assumptions, (ii) the optimisation of the energy supply for stand-alone direct air capture systems, and (iii) the endogenous determination of electricity prices by embedding direct air capture systems within an integrated energy system model. The three methodologies have different strengths and weaknesses. The first is straightforward and transparent, but contains a methodological simplification as electricity prices influence DACCS investment – and DACCS investment influences electricity prices. The second methodology, stand-alone optimisation, determines the cost of electricity endogenously, but the repercussions of the existing electricity system's operation are not part of this approach. It seems likely that the cost for electricity is overestimated in such a set-up. For example, the remaining electricity system provides flexibility that can also be utilised by the DACCS system. DACCS can use electricity at times when RES generation is otherwise curtailed in the system, especially considering that large-scale DACCS, as a technology to compensate remaining emissions in hard-to-abate sectors, is likely to be built when the electricity system is already based on renewable energy to a large extent. On the other end of the price range, DACCS are likely to cease production during periods wherein capacity is scarce and prices are spiking. These effects will lower the average electricity price for DACCS operation. The third methodology, embedding DACCS in existing systems, is likely to be the most accurate. At the same time, it is also the most complex approach, as an energy system model is required.

### 2.2.1 Exogenous Electricity Prices

A detailed sampling of the literature with regard to (i), i.e. exogenous electricity prices, revealed several relevant publications. An analysis of three direct air capture processes (two liquid sorbent and one solid sorbent) was conducted by Sabatino et al. (2021). They found that varying electricity prices between 20 and 100 EUR/MWh could lead to total DACCS cost of less than 200 USD/t$CO_2$ in all three configurations. The authors considered the energy demand, electricity and heat prices, productivity and air contactor prices, but not storage and transportation as part of the DACCS cost.

Sendi et al. (2022) conducted a global analysis to investigate the impact of regional climate variations on the performance of direct air capture systems. Testing three different electricity prices of 0, 50, and 100 USD/MWh, they found the capturing cost to be between 220 and 565 USD/t$CO_2$ for favourable cold and dry climate conditions.

By combining an engineering model with technological learning projections, Young et al. (2023) estimated cost trajectories for DACCS until 2050 across seven different countries. They found that the cost could range between 100 and 600 USD/t$CO_2$. As the study used a top-down cost evaluation approach, electricity sources and prices were assumed as exogenous parameters for the specific countries accounting for solar PV and wind energy as the only supply sources, resulting in assumed prices that vary between 5 and 20 USD/MWh.

Sievert et al. (2024) projected future direct air capturing costs by applying experience rates of mature technologies to three direct air capture technologies. By simulating scaling of capturing capacity to up to 1 Gt$CO_2$/year they found the DACCS cost to be between 281 and 579 USD/t$CO_2$. The authors assumed the energy supply to be via solar PV and battery storage, wind, nuclear, or geothermal, leading to price assumptions that varied between 17 USD/MWh and 71 USD/MWh

Terlouw et al. (2024) presented a geospatial analysis of the economic performance of DACCS across Europe, accounting for the impact of location-specific weather conditions on the energy demand of the direct air capture units. The electricity demand was supplied from either only the grid or from curtailed renewable generation with predetermined prices for each country, varying between 42 EUR/MWh and 133 EUR/MWh.

Mühlbauer et al. (2025) provided techno-economic insights for integrating DACCS into future energy and climate strategies, supporting its role in large-scale carbon removal by 2100. With an electricity price of around 17 EUR/MWh, they estimated costs below 100 EUR/t$CO_2$ by 2050. However, the study did not account for regional variations in direct air capture system performance and location-specific constraints, like the electricity grid, thus neglecting the impact on the OPEX.

In summary, the prices assumed in the literature vary widely. They range from 5 EUR/MWh, as reported by Young et al. (2023), to 133 EUR/ MWh, as reported by Terlouw et al. (2024).

### 2.2.2 Endogenous Optimisation for DACCS in Isolation

Another branch of research on DACCS models isolated energy systems in which DACCS is the sole consumer, leading to large spreads in estimated operating costs between 50 and 600 EUR/t$CO_2$ (e.g. Fasihi et al. 2019; Young et al. 2023; Terlouw et al. 2024; Wenzel et al. 2025a). This methodology accounts for energy costs endogenously by optimising investment in and dispatch of generation technologies needed to meet the electricity demand of DACCS.

In 2019, Fasihi et al. reviewed state-of-the-art direct air capture technologies and conducted a techno-economic analyses of direct air capture systems based in Morocco. They estimated the $CO_2$ capture cost of an LT direct air capture system powered by solar PV, wind, and batteries to range between 32 and 70 EUR/t$CO_2$ for 2050.

Wiegner et al. (2022) conducted an analysis that considered next to exogenous electricity prices as an optimisation of the energy supply of a stand-alone direct air capture plant using solar PV, wind, and battery technology as possible investment candidates. They concluded that direct air capture systems should be assessed along with the energy supply system, stating that the available and cheap renewable supply influences the optimal system configuration, outweighing even bad direct air capture site conditions.

In a more recent case study, Wenzel et al. (2025a) investigated the effects of weather variability on the carbon capture cost of direct air capture systems powered by a different combination of wind, solar PV, and battery technologies in Germany. Due to the significant influence of humidity and air temperature on the electrical and thermal energy requirements, capture costs in a 2045 scenario were found to vary widely. For LT direct air capture systems, costs ranged from 223 to 848 EUR/t$CO_2$, with an average of 285 EUR/t$CO_2$ excluding $CO_2$ transport and storage where calculated.

The drawback of stand-alone approaches is that they neglect the inherent flexibility of electricity systems, which drives up electricity costs. Island solutions tend to have higher overall supply costs (Kaundinya et al. 2009; Ortega-Arriaga et al. 2021; Basnet et al. 2023).

### 2.2.3 Embedding DACCS in Integrated Energy System Models

A few studies integrate direct air capture into broader electricity system models, which has the advantage that the resulting energy costs for the system are not only endogenously quantified but also take into account repercussions with other system components, on both the supply and demand sides. However, these studies neglect the underlying electricity grid or assume uniform storage and transport cost per ton of $CO_2$, overlooking the geospatial scale, impacting the electricity and $CO_2$ transport and storage infrastructure; this significantly affects the DACCS cost.

Breyer et al. (2020) conducted a case study on the deployment of direct air capture systems in the Maghreb region, integrating them into the “LUT Energy System Transition Model”. They found that a PV-dominated system, supplemented by wind and battery storage, enabled near-baseload operation of direct air capture units leading to an estimated DACCS cost of approximately 55 EUR/t$CO_2$ by 2050.

Lehtveer and Emanuelsson (2021) conducted a case study for two regions in Europe by integrating direct air capture and BECCS systems into an energy system model for the year 2050. Although BECCS had a lower capture cost than DACCS, the overall system costs were lower with DACCS integration. This is because the levelized cost of electricity generation from BECCS is higher, while the DACCS system uses cheaper renewable electricity and is more flexible. They calculated the DACCS cost to be between 105 and 55 EUR/t$CO_2$ and concluded that the metric of carbon capture cost has limited explanatory power when it comes to assessing the economic efficiency of direct air capture systems.

In 2023 Lux et al. built on the "Enertile" model to investigate the impacts of the integration of DACCS into a carbon-neutral European energy system 2050. Using different techno-economic cost assumptions, they estimated DACCS costs between 160 and 270 EUR/t$CO_2$ under conservative assumptions, and 60 and 140 EUR/t$CO_2$ with more optimistic parameters. Their approach accounted for the energy system impacts on the DACCS cost, showing that electricity costs dominate.

## 2.3 $CO_2$ Transport and Storage Costs of DACCS

The deployment of $CO_2$ storage infrastructure remains politically contested in many countries, particularly regarding whether storage sites should be located onshore or offshore. Public perceptions vary considerably across countries, with populations in countries such as Norway generally expressing more positive attitudes, while acceptance tends to be lower in countries such as Germany. Empirical studies suggest that the distinction between onshore and offshore storage does not substantially alter overall public acceptance, as both options face relatively low levels of support. Nevertheless, offshore storage is often perceived somewhat more favourably, primarily because storage sites are located farther away from populated areas (Schumann et al. 2014; Chan et al. 2024). Despite these societal concerns, onshore storage sites are increasingly discussed as an important component of future $CO_2$ management strategies (Allen, 2025). Although the estimated $CO_2$ transport and storage costs typically account for only around 10 % of total DACCS costs, they remain non-negligible.

Consequently, understanding whether permitting onshore storage could significantly reduce overall system costs is an important policy-relevant question.

In most studies, the cost for transport and storage is considered exogenous and is added to the direct air capturing cost. Thus e.g. Lux et al. (2023) assumed a uniform sequestration cost of 10 EUR/t$CO_2$ for each country considered. Young et al. (2023) assumed an additional storage cost of 11 USD/t$CO_2$, and a transportation cost between 0.02 and 0.048 UDS/t$CO_2$. Sievert et al. (2024) included the cost as part of annualised variable OPEX with a transportation cost of 0.09 USD/t$CO_2$ and a storage cost between 11 and 27 UDS/t$CO_2$. Mühlbauer et al. (2025) differentiated the transport costs of an onshore pipeline to between 0.077 and 0.104 UDS per t$CO_2$/km, and two different onshore and offshore storage types with costs varying between approximately 2.5 and 11 USD per t$CO_2$/km.

A more detailed estimate of the transportation cost was carried out by Terlouw et al. (2024). The authors considered the additional transportation cost for location-specific direct air capture units depending on the shortest path to the closest storage site. Applied to all of Europe, this approach accounts more accurately for the location-specific transportation cost, which can be significant when the direct air capture unit is located far away from the nearest storage site. The transport cost is 0.04 EUR per t$CO_2$/km, and the uniform storage cost is 11 EUR/t$CO_2$, assuming no storage capacity restrictions.

Taken together, we find that in the literature the $CO_2$ transport and storage costs account for between 4 and 16 % of the total long-term (2050) DACCS cost estimates. Although the $CO_2$ transportation and storage costs are found to be less significant to the total DACCS cost compared to the direct air capture units CAPEX and energy cost, the omitted infrastructure constraints may lead to an incorrect performance assessment when direct air capture siting decisions must be made. In general, the representation of the infrastructure is often simplified, leaving aspects such as the joint usage of infrastructure or the differentiation between onshore and offshore storage expansion unaddressed. To the best of our knowledge, no study couples electricity and $CO_2$ infrastructure expansion, representing endogenous dependencies, while estimating DACCS cost.

## 2.4 Research Gap and Contribution

As the literature review reveals, there are several research gaps remaining when it comes to the assessment of future DACCS cost and the systemic impacts of its infrastructure expansion.

- First, we observed a wide range in electricity price estimates, derived in the existing literature from three different methodological approaches. There is no investigation into how these different methodological approaches affect the resulting capture cost estimates of the DACCS process. We address this issue by analysing and comparing three distinct modelling approaches, assessing their influence on the resulting DACCS and electricity system cost, and examining their impact on the generation technology mix.
- Second, there is a controversial debate in regards to where to store the $CO_2$. However, the effect of restricting $CO_2$ storage to offshore sites only versus allowing additional national or local onshore storage options on the electricity and $CO_2$ infrastructure remains largely unexplored. In this study, we therefore consider two different $CO_2$ storage location scenarios: North Sea offshore storage only and allowing onshore storage sites to be expanded.
- Third, existing energy system modelling studies often have limited spatial resolution. As grid capacities pose binding constraints in many electricity systems, RES potentials are geographically dispersed, the segregated $CO_2$ requires transport and storage, which depends on storage availability, and studies with higher spatial resolution are lacking but would be valuable. In addition, local temperature and humidity effects – which significantly influence the direct air capture system's energy requirements and efficiency – are often simplified considerably. This study addresses these gaps by integrating DACCS into a high-resolution European electricity generation and transmission expansion model for the year 2050.
- Fourth, the potentially significant contribution of DACCS in a carbon-neutral economy crucially depends on its cost competitiveness. While the large variations in cost estimates in the literature are reasonable and justified by the uncertainty of a new technology, there has not yet been a study that explores the different influences that various assumptions regarding electricity prices and storage sites have on the capture cost. By combining spatially-explicit modelling with techno-economic and system-level insights, our study provides a more realistic and policy-relevant understanding of a potential DACCS deployment and its cost in Europe in 2050.

# 3 MODEL, SET-UP AND DATA, AND SCENARIOS

This section first introduces the nomenclature for the mathematical formulation of the optimisation model, which is described in section 3.1. Second, section 3.2 presents the empirical use case of the underlying study. Third, section 3.3 describes the investigated scenarios.

## Nomenclature

**Indices and Sets**

| | |
|---|---|
| $t \in T$ | Timesteps |
| $n \in N$ | Network nodes, (alias $nn$) |
| $l \in L$ | Network transmission lines |
| $s(l) \in N$ | Sending (from/origin) node of line l |
| $r(l) \in N$ | Receiving (from/origin) node of line l |
| $c \in C$ | Cycle in network |
| $i \in I$ | Electricity generation technologies |
| $st \in ST$ | Storage technologies |
| $p \in P$ | $CO_2$ pipelines |
| $s(p) \in N$ | Sending (from/origin) node of pipe p |
| $r(p) \in N$ | Receiving (from/origin) node of pipe p |
| $acl \in ACL \subseteq L$ | Alternating current lines |
| $dcl \in DCL \subseteq L$ | Direct current lines |
| $r \in R \subseteq I$ | Weather dependent generation units |
| $d \in D \subseteq I$ | Dispatchable generation units |
| $hy \in HY \subseteq D$ | Hydro generation units |
| $ror \in HY$ | Run of River power plants |
| $rsv \in HY$ | Reservoir power plants |
| $b \in ST$ | Battery storage units |
| $psp \in ST$ | Pumping storage power plant |
| $hst \in ST$ | Hydrogen storage |

**Parameters**

| | |
|---|---|
| $ac_i$ | Annualized investment costs for generation units [EUR/MW] |
| $ac_{st}^{STOR}$ | Annualized investment costs for storage capacity [EUR/MWh] |
| $ac_{st}^{CH}$ | Annualized investment costs for storage charging power [EUR/MW] |
| $ac_{st}^{DIS}$ | Annualized investment costs for storage discharging power [EUR/MW] |
| $ac_i^{Ex}$ | Annualized costs existing generation capacity [EUR/MW] |
| $ac^{HP}$ | Annualized investment costs for heat pumps [EUR/$MW_{th}$] |
| $ac^{DAC}$ | Annualized investment costs for direct air capture units [EUR/$tCO_2$/h] |
| $ac^{CS}$ | Annualized investment costs for $CO_2$ storage units [EUR/$tCO_2$] |
| $ac^{L}$ | Annualized investment costs for transmission line [EUR/MW] |
| $ac^{PIP}$ | Annualized investment costs for $CO_2$ pipeline [EUR/$tCO_2$/h] |
| $vc_{i,t}$ | Variable electricity generation cost [EUR/MWh] |
| $sc$ | Load shedding costs [EUR/MWh] |
| $heat_{n,t}^{DAC}$ | Locational & time dependent heat demand of direct air capture [$MWh_{th}$/$tCO_2$] |
| $el^{CS}$ | Electricity demand $CO_2$ storage [MWh/$tCO_2$] |
| $cop_{n,t}$ | Coefficient of performance from heat pump |
| $cf_{i,t}$ | Technology & time dependent capacity factor |
| $cap_{n,i}^{EX}$ | Existing electricity generation capacity [MW] |
| $cap_{n,st}^{EXDIS}$ | Existing storage discharging capacity [MW] |
| $cap_{n,st}^{EXCH}$ | Existing storage charging capacity [MW] |
| $cap_{n,st}^{EXSTOR}$ | Existing storage capacity [MWh] |
| $\sigma_{st}^{CH}$ | Storage charging efficiency [%] |
| $\sigma_{st}^{DIS}$ | Storage discharging efficiency [%] |
| $lvl_{n,st,t=1}^{start}$ | Start filling level of storages [MWh] |
| $cm_{c,acl}$ | Cycle Matrix for AC lines |
| $x_{acl}$ | Reactance of AC line [Ω] |
| $cap_l^{EXL}$ | Existing transmission line capacity [MW] |

**Positive Variables**

| | |
|---|---|
| TC | Annual total system cost [EUR] |
| $CAP_{n,i}$ | Capacity of generation technology [MW] |
| $CAP_{n,st}^{STOR}$ | Storage capacity [MWh] |
| $CAP_{n,st}^{CH}$ | Storage charging capacity [MW] |
| $CAP_{n,st}^{DIS}$ | Storage discharging capacity [MW] |
| $CAP_n^{HP}$ | Capacity of heat pump [MW] |
| $CAP_n^{DAC}$ | Capacity of direct air capture [$tCO_2$/h] |
| $CAP_n^{CS}$ | Capacity of $CO_2$ storage [$tCO_2$] |
| $CAP_l^{L}$ | Transmission line capacity [MW] |
| $CAP_p^{PIP}$ | $CO_2$ pipeline capacity [$tCO_2$/h] |
| $GEN_{n,i,t}$ | Electricity generation of generator [MWh] |
| $DIS_{n,st,t}^{STOR}$ | Discharging of electricity storage [MWh] |
| $CH_{n,st,t}^{STOR}$ | Charging of electricity storage [MWh] |
| $LVL_{n,st,t}^{STOR}$ | Storage filling level [MWh] |
| $LS_{n,t}$ | Load shedding [MWh] |
| $GEN_{n,t}^{DAC}$ | Capturing process of $CO_2$ direct air capture unit [$tCO_2$] |

| | | | |
|---|---|---|---|
| $vom^{CS}$ | Variable operation and maintenance cost of $CO_2$-storage [EUR/$tCO_2$] | $GEN_{n,t}^{HP}$ | Generation of heat pump unit [$MWh_{th}$] |
| $dem_{n,t}^{EL}$ | Nodal electricity demand [MWh] | $CH_{n,t}^{CS}$ | $CO_2$ compression of storage unit [$tCO_2$/h] |
| | | **Free Variables** | |
| $el_{n,t}^{DAC}$ | Locational & time dependent electricity demand of direct air capture [MWh/$tCO_2$] | $F_{l,t}^{L}$ | Power flow on transmission line [MWh] |
| | | $F_{p,t}^{PIP}$ | $CO_2$ flow on pipeline [$tCO_2$/h] |

## 3.1 Mathematical Model Formulation

We develop a bottom-up linear optimisation model of a climate-neutral electricity system to analyse DACCS costs under three different modelling approaches. First, direct air capture systems are modelled in stand-alone operation, where electricity supply is built exclusively to serve their energy demand. Second, they are integrated into an existing electricity system. The third approach is a joint optimisation of the power system and DACCS deployment. All three model set-ups are combined with two different $CO_2$ storage site assumptions (North Sea offshore only, and storage distributed across Europe). In each set-up, the model identifies the cost-optimal capacity and transmission expansion options by jointly optimising the power mix and the deployment of the $CO_2$ infrastructure. The model is formulated as follows:

$$\textbf{Min}: \quad TC = \sum_{n\in N}\Bigg(\sum_{i\in I} CAP_{n,i}\cdot ac_i + \sum_{st\in ST}(CAP_{n,st}^{STOR}\cdot ac_{st}^{STOR} + CAP_{n,st}^{CH}\cdot ac_{st}^{CH} + CAP_{n,st}^{DIS}\cdot ac_{st}^{DIS}) + \sum_{i\in I} cap_{n,i}^{EX} \cdot ac_i^{Ex} + \sum_{i\in I, t\in T} GEN_{n,i,t}\cdot vc_{i,t} + \sum_{t\in T} LS_{n,t}\cdot sc + CAP_n^{HP}\cdot ac^{HP} + CAP_n^{DAC}\cdot ac^{DAC} + CAP_n^{CS}\cdot ac^{CS} + \sum_{t\in T} CH_{n,t}^{CS}\cdot vom^{CS}\Bigg) + \sum_{l\in L} CAP_l^{L}\cdot ac^{L} + \sum_{p\in P} CAP_p^{PIP}\cdot ac^{PIP} \tag{1}$$

**subject to:**

$$dem_{n,t}^{EL} = \sum_{i\in I} Gen_{n,i,t} + \sum_{st\in ST}\left(DIS_{n,st,t}^{STOR} - CH_{n,st,t}^{STOR}\right) + \sum_{l\in r(l)} F_{l,t}^{L} - \sum_{l\in s(l)} F_{l,t}^{L} - GEN_{n,t}^{DAC}\cdot el_{n,t}^{DAC} - \frac{GEN_{n,t}^{HP}}{cop_{n,t}} - CH_{n,t}^{CS}\cdot el^{CS} - LS_{n,t} \qquad \forall\, n,t \tag{2}$$

$$Gen_{n,i,t} \le (cap_{n,i}^{EX} + CAP_{n,i})\cdot cf_{i,t} \qquad \forall\, n,i,t \tag{3}$$

$$DIS_{n,st,t}^{STOR} \le cap_{n,st}^{EXDIS} + CAP_{n,st}^{DIS} \qquad \forall\, n,st,t \tag{4}$$

$$CH_{n,st,t}^{STOR} \leq cap_{n,st}^{EXCH} + CAP_{n,st}^{CH} \qquad \forall\, n, st, t \tag{5}$$

$$LVL_{n,st,t}^{STOR} \leq cap_{n,st}^{EXSTOR} + CAP_{n,st}^{STOR} \qquad \forall\, n, st, t \tag{6}$$

$$LVL_{n,st,t}^{STOR} = LVL_{n,st,t-1}^{STOR} + \left( CH_{n,st,t}^{STOR} \cdot \sigma_{st}^{CH} \right) - \frac{DIS_{n,st,t}^{STOR}}{\sigma_{st}^{DIS}} \qquad \forall\, n, st, \forall t \tag{7}$$

$$\sum_{acl \in ACL} CM_{c,acl} \cdot F_{acl,t}^{L} \cdot x_{acl} = 0 \qquad \forall\, c, t \tag{8}$$

$$-\left(cap_{l}^{EXL} + CAP_{l}^{L}\right) \leq F_{l,t}^{L} \leq cap_{l}^{EXL} + CAP_{l}^{L} \qquad \forall\, l, t \tag{9}$$

$$LS_{n,t} \leq dem_{n,t}^{EL} \qquad \forall\, n, t \tag{10}$$

$$GEN_{n,t}^{DAC} + \sum_{p \in r(p)} F_{p,t}^{PIP} = CH_{n,t}^{CS} + \sum_{p \in s(p)} F_{p,t}^{PIP} \qquad \forall\, n, t \tag{11}$$

$$GEN_{n,t}^{HP} \leq CAP_{n}^{HP} \qquad \forall\, n, t \tag{12}$$

$$GEN_{n,t}^{DAC} \leq CAP_{n,t}^{DAC} \qquad \forall\, n, t \tag{13}$$

$$GEN_{n,t}^{DAC} \cdot heat_{n,t}^{DAC} \leq GEN_{n,t}^{HP} \qquad \forall\, n, t \tag{14}$$

$$F_{p,t}^{PIP} \leq CAP_{p}^{PIP} \qquad \forall\, p, t \tag{15}$$

The objective function in eq. (1) minimises the total system cost, which is the sum of the annualised cost ($ac$) from investments into new electricity generation, storage units and transmission lines, the cost of the existing power plants, the variable generation ($vc$), and load shedding cost ($sc$). Furthermore, investment costs of heat pumps, direct air capture units, $CO_2$ pipelines and storage, and the variable operation and maintenance costs of $CO_2$ storage units ($vom$) are included. The energy balance constraint is stated by eq. (2), which ensures that the electricity demand for each node is equal to all electricity generation, storage charging and discharging actions, power flows, electricity consumption from direct air capture units, heat pumps, and $CO_2$ storage, and load shedding actions at all timesteps. The electricity generation is constrained by the sum of the existing capacity and new installations related to the capacity factor for the specific generation technology in eq. (3). The storage charging and discharging power, as well as the maximum storage capacity, are constrained by the existing and invested capacities as stated by eqs. (4), (5), and (6). Eq. (7) determines the storage-filling level for all timesteps. The linear optimal power flow constraint is defined using the cycle flow formulation as stated by Neumann et al. (2022) in eq. (8), and the line capacity restriction is defined by eq. (9). Eq. (10) constrains the maximum possible load shedding. The nodal $CO_2$ balance constraint in eq. (11) ensures that the amount of $CO_2$ capturing, storing, and flow is equal for each time step. The maximum heat generation of the heat pump is restricted by eq. (12), and the maximum absorption of the direct air capture unit by eq. (13). Furthermore, the absorption heat demand of the direct air capture unit must be provided by the heat generation though the heat pump as stated in eq. (14). Finally, eq. (15) restricts the upper $CO_2$ flow limit though the installed pipeline capacity.

## 3.2 Empirical Use Case

Our study encompasses 32 European countries with an underlying network of 128 electricity nodes and 273 lines. We derived the network and the technological data of the transmission lines using the algorithm from PyPSA (2024) to cluster the underlying ENTSO-E transmission system data. We supplemented 12 additional nodes to include the offshore storage locations. An overview of the system, including the considered onshore and offshore $CO_2$ storage locations, is presented in Figure 1. For each electricity node in the system, a timeseries for the electricity demand, capacity factors, temperature, and humidity are provided in two-hourly resolution. The expansion model optimises investment in and dispatch of electricity generation technologies, transmission lines, direct air capture system deployment, $CO_2$ transport, and storage. The carbon-neutral electricity sector comprises nuclear, biomass, solar PV, onshore and offshore wind, battery storage, $H_2$-fired OCGT, electrolysers, hydrogen storages, hydro power such as run of river, reservoir, and pumped storage, as well as heat pumps and transmission lines (both AC and DC). Investment is freely-optimised, except for nuclear, biomass, and hydro power, which are fixed spatially accumulated to their currently (2025) installed capacities. The $CO_2$ sector also includes expandable low-temperature direct air capture systems, $CO_2$ pipelines, and geological $CO_2$ onshore and offshore storage sites (i.e. deep saline formations and depleted hydrocarbon fields). We set the yearly capturing target in our case study to 100 Mt (megatons) $CO_2$ per year. A detailed overview of the specific parameter and cost assumptions of the electricity generation, direct air capture units, $CO_2$ pipeline, and storage technologies is provided in Appendix A.

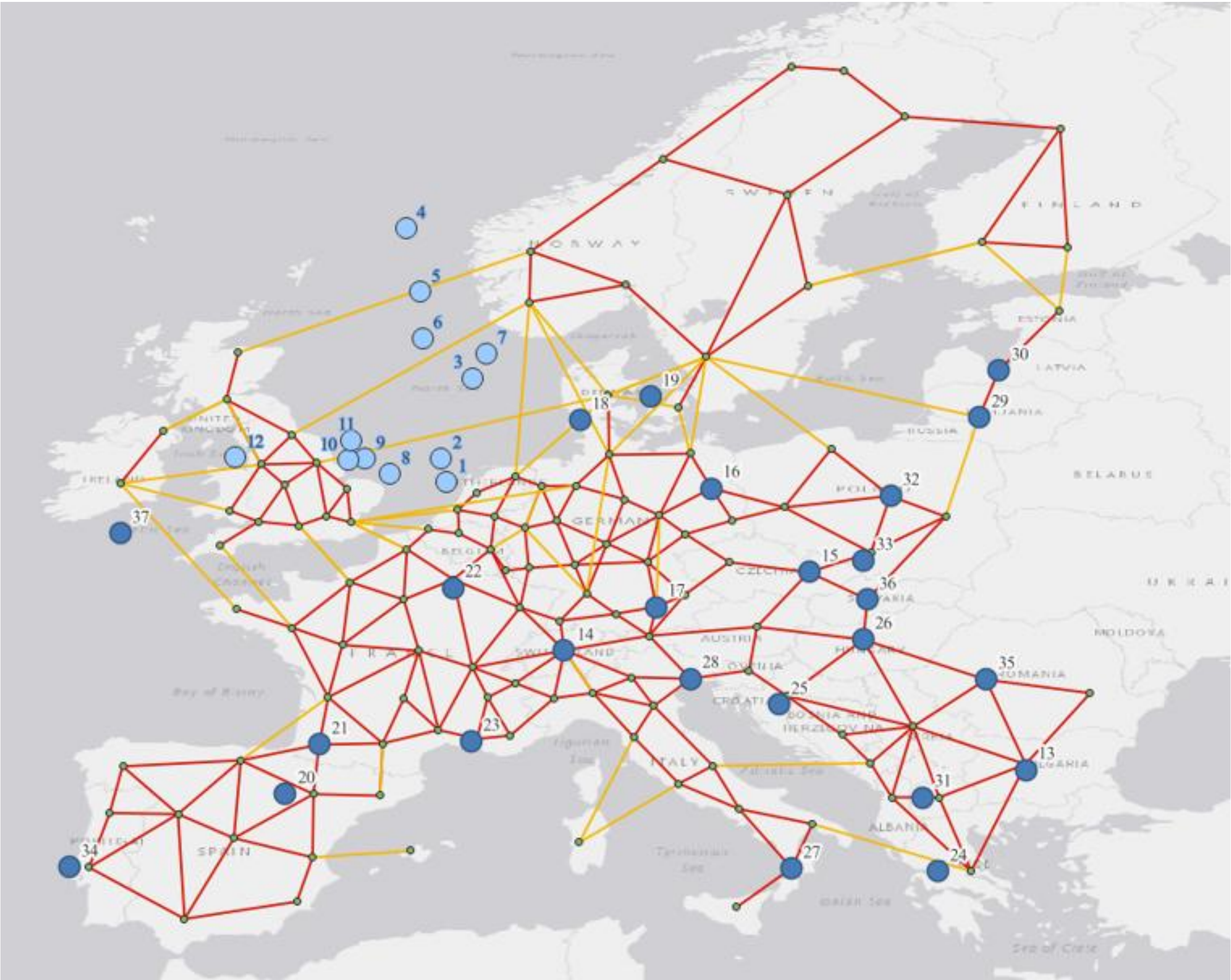

**Figure 1: Electricity system with numbered $CO_2$ storage locations, with individual site information provided in Appendix A. The large light blue nodes correspond to the North Sea and UK and dark blue nodes to other onshore and offshore storage projects and potentials. The remaining small green dots are nodes in the electricity system with the AC represented by red and the DC by yellow lines. Data Source: CATF, PCI-PMI from the European Commission.**

## 3.3 Scenarios

We first compute a "Benchmark" scenario, which optimises a carbon-neutral European electricity system without any DACCS activity. The resulting capacity and generation mix is presented in Appendix B. Furthermore, we model six additional scenarios, where an identical amount of 100 Mt $CO_2$ needs to be captured, transported, and stored with DACCS systems. The six scenarios differ in two dimensions: the "Modelling Approach" on the calculation of the electricity price and the "Carbon Storage Location", as visualised by Table 1 below.

**Table 1: CCS Scenario Overview**

| Modelling Approach \ Carbon Storage Location | North Sea | Europe |
|---|---|---|
| Isolated | Isolated North Sea (IsNS) | Isolated Europe (IsEU) |
| Added | Added North Sea (AdNS) | Added Europe (AdEU) |
| Integrated | Integrated North Sea (IgNS) | Integrated Europe (IgEU) |

The first dimension, the modelling approach, differentiates the electricity system under consideration, which varies between "Isolated", "Added", and "Integrated".

- Isolated: Electricity generation capacity is modelled solely to supply the DACCS activity, allowing no interaction with the existing electricity system. DACCS is thus the only consumer in the system; all capacity is exclusively built to satisfy that demand; all costs are thus attributed to DACCS.
- Added: Electricity demand from DACCS activity is added to an existing system. The existing system consists of the Benchmark scenario's capacities. The model in the Added approach takes these capacities as exogenous, but is free to add new generation capacity to meet the electricity demand from DACCS, which is added to the non-DACCS demand. The cost difference between the cost in this added DACCS scenario and the cost in the benchmark scenario are attributed to DACCS.
- Integrated: This is an approach in which only the hydro, nuclear, biomass, and transmission line capacities are added as exogenous parameters. The model can optimise all other generation capacities and transmission lines to meet the total demand, both from the European electricity systems and from DACCS activity. Thus, the DACCS electricity demand is integrated from the start, allowing the model to co-optimise the generation and transmission capacities across Europe holistically.

The second dimension, "Carbon Storage Location", determines where storage of $CO_2$ is allowed. $CO_2$ can be stored either in the North Sea only or in numerous potential carbon storage locations throughout Europe (as visualised in Figure 1). This variation investigates how geographic storage constraints, transportation, and the system integration of DAC units influence the cost of negative $CO_2$ emissions and the design of the energy system.

# 4 RESULTS

Our results analyse the expansion of the $CO_2$ infrastructure for each of the six CCS scenarios in section 4.1. Second, we examine the impacts on the optimal scenario-specific electricity supply for the direct air capture units in section 4.2. Finally, we analyse the individual DACCS cost components and the resulting costs in section 4.3.

## 4.1 $CO_2$ Infrastructure Expansion

Figure 2 presents the optimal expansion of the $CO_2$ infrastructure in our six scenarios. The six image segments are divided into three rows to illustrate the effect of each modelling approach (Isolated, Added, or Integrated) and two columns to show the impact of storage locations (North Sea only versus various European locations) on the expansion decisions. The figure shows the capacities of direct air capture units (green dots, in Mt of $CO_2$ captured per year), the carbon transport vectors (in black), and the underground storage (grey bars, in Mt $CO_2$ per year). Finally, the total direct air capture capacity (Mt per year) is depicted in the top right, and the scenario-specific abbreviations are in the lower left corner of each individual segment.

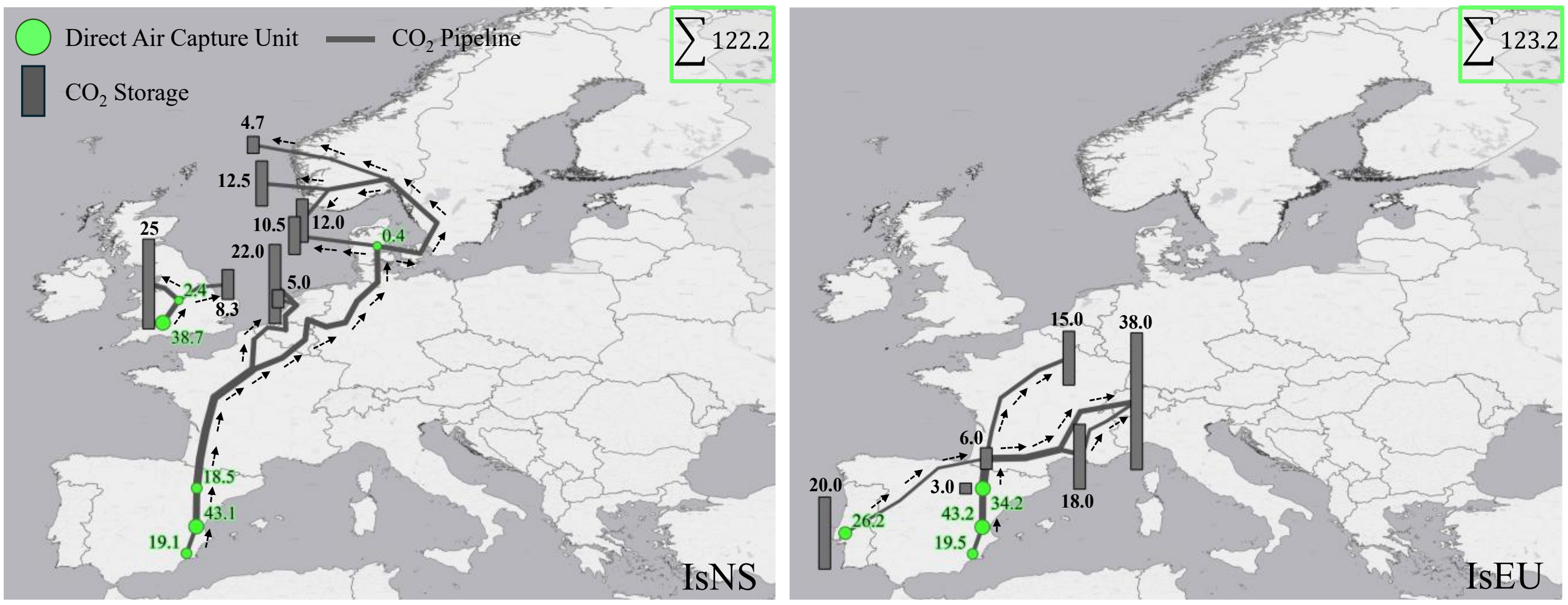

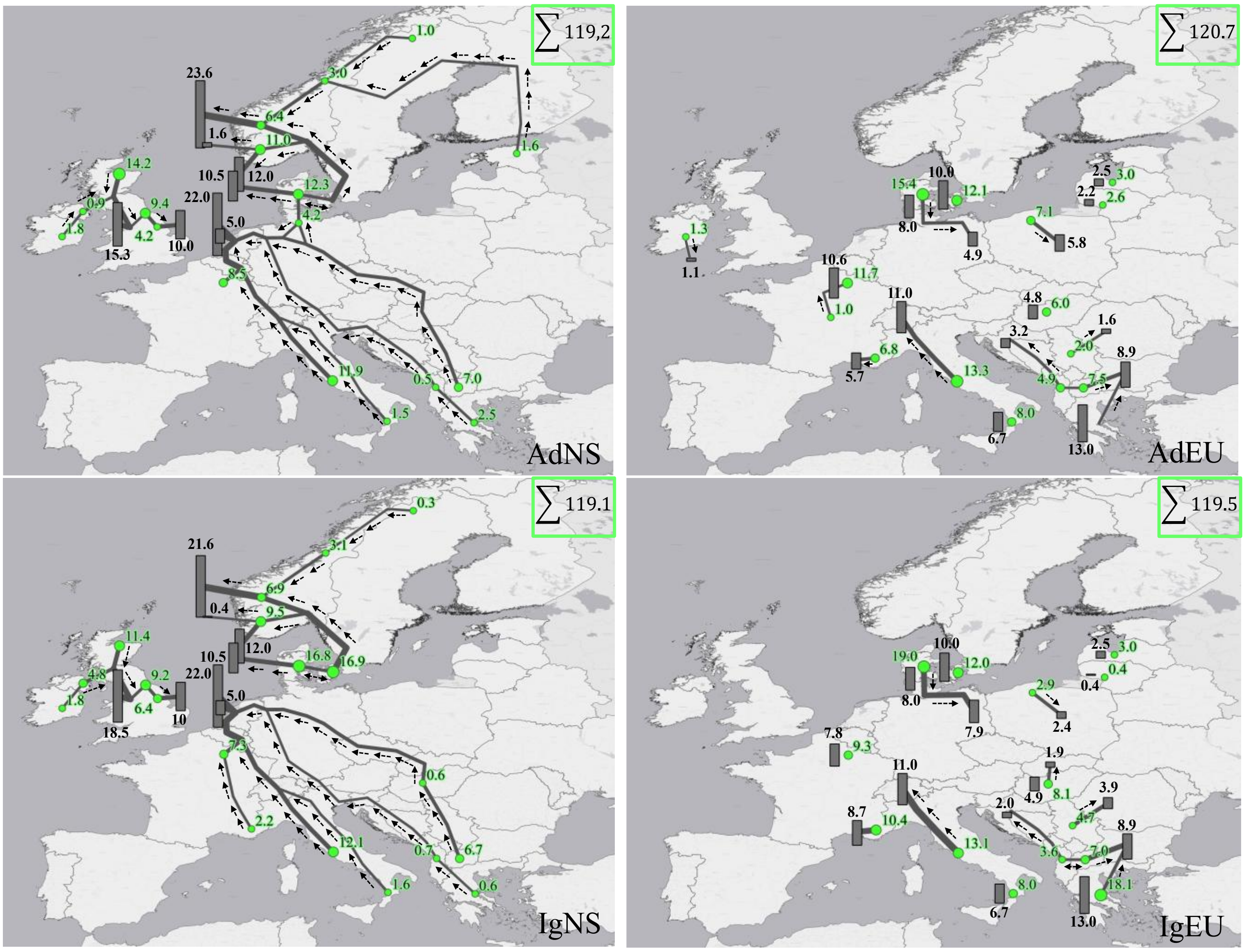

**Figure 2: Scenario-specific expansion of DACCS-related infrastructure**

The results differ significantly between scenarios. Comparing modelling approaches, the Isolated scenarios (top row) place direct air capture units mostly in south-western Europe to tap into the high-quality RES supply. The other four scenarios, all allowing higher integration with the existing electricity infrastructure, distribute direct air capture systems over a wider range of locations, because installed electricity generation capacity can be partially utilised.

Furthermore, the Isolated scenarios result in the highest installed direct air capture capacities (numbers in top-right corner of each panel). The Added and Integrated scenarios have higher utilisation rates for the direct air capture systems as a more constant low-cost electricity baseload supply is available. Hence, they require less capacity.

A detailed discussion of the individual differences and the impact of storage siting (left vs. right panels) can be found in Appendix C. In the following section, we examine the underlying energy system dynamics driving these outcomes in greater detail.

## 4.2 Electricity Generation Capacity Additions

As mentioned in the previous section, the optimal integration of DACCS in electricity systems has different components: the placement of the direct air capture unit (electricity consumption), the placement of the additional electricity generation capacity required for that (electricity supply), the placement of the $CO_2$-sink (electricity consumption), and the resulting transportation of both electricity and $CO_2$. For the electricity consumption of the direct air capture units, warmer countries with lower humidity require less thermal energy but more electricity. Since the thermal energy demand is supplied by a heat pump, the system's total energy requirement can be expressed entirely in terms of electricity demand. Although higher temperatures increase the electricity consumption of the DAC unit, the reduction in thermal energy demand generally dominates, as heat provision accounts for roughly 60-70 % of total electricity use. Consequently, lower heat demand typically leads to a reduction in overall electricity consumption.

Figure 3 below shows the (additional) electricity capacity installed to supply the DACCS electricity demand for each scenario. We compute the additional capacity from the Added and Integrated scenarios as the difference from the Benchmark scenario without the negative emissions target.

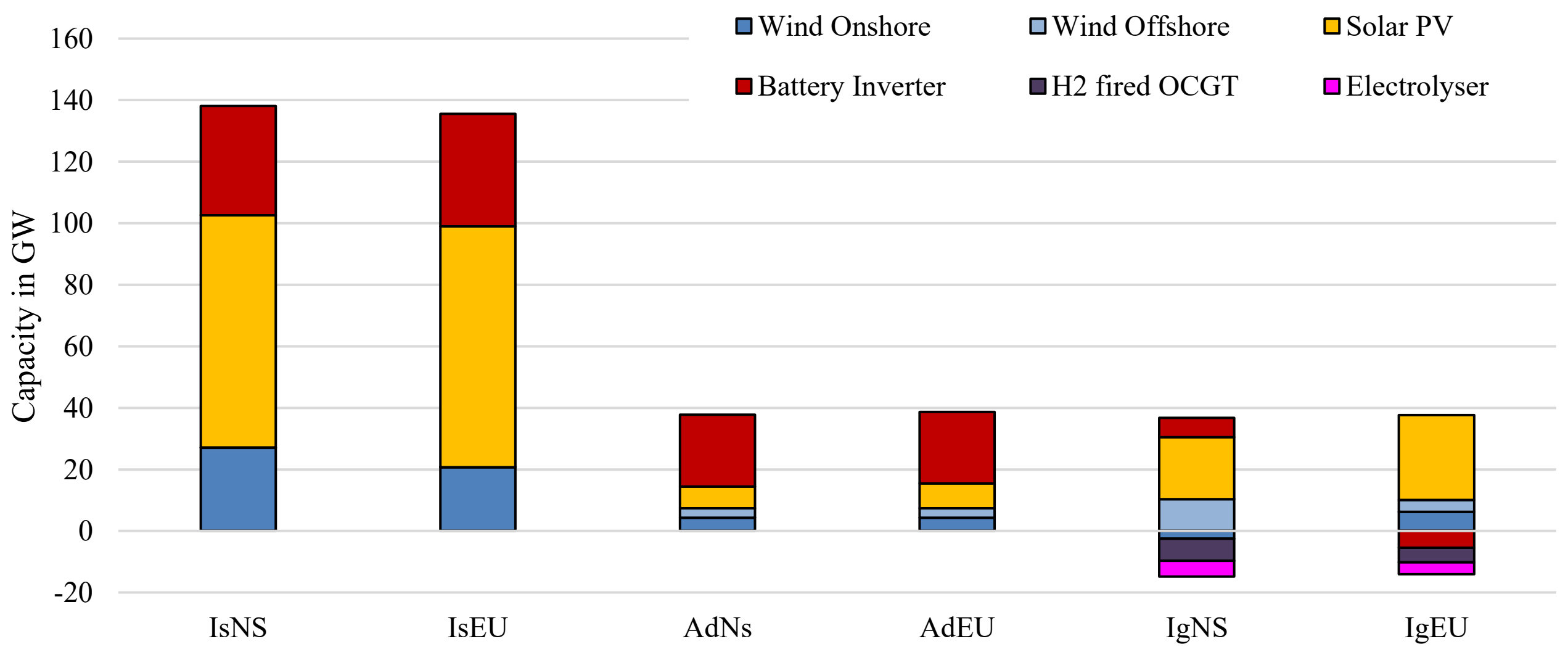

**Figure 3: Additional capacity to supply the direct air capture and $CO_2$ storage electricity demand for each scenario**

Two main findings emerge. First, the large difference in installed capacity between the two Isolated scenarios on the one hand and the four more integrated scenarios on the other hand. While a certain difference was to be expected because no existing infrastructure is available and all demand must be met by newly built generators in the Isolated scenarios, the Isolated scenarios require more than three times more capacity (from <40 GW of additional capacity in the four more integrated scenarios to nearly 140 GW in the two Isolated scenarios). This result confirms the high synergies between DACCS and the main part of the electricity system. The key synergy, which lies in a higher utilisation of renewable energy, i.e. reduced curtailment, will be analysed in detail after we discuss the specific technologies in which the model invests. This is the second major finding in Figure 3. Most of the additional capacity is solar PV, complemented with battery storage. In the Isolated scenarios, about 75 and 78 GW of solar PV are installed, supported by roughly 35 GW of battery inverter capacity and additional onshore wind (27 GW in IsNS and 20 GW in IsEU). In the Added scenarios, 7-8 GW of solar PV are complemented with around 23 GW of batteries and around 7 GW from onshore and offshore wind. The two Integrated scenarios also add mostly solar power, complemented by wind offshore and batteries in IgNS, and both on- and offshore wind power in IgEU. Electrolyser and H2-fired OCGT capacities decrease in both Integrated scenarios, but the difference is small.

Further insights can be gained by looking at generation instead of capacity. Figure 4 below depicts the additional electricity generation to supply the direct air captures electricity demand for each scenario. We calculate the additional generation from the Added and Integrated scenarios as the difference from the Benchmark scenario without the negative emissions target. The upper left corner of each graph shows the total amount of electricity consumed for DACCS. Note that despite the vast differences in installed capacity just discussed, the numbers are of the same order of magnitude (between 160,960 and 182,060 GWh).

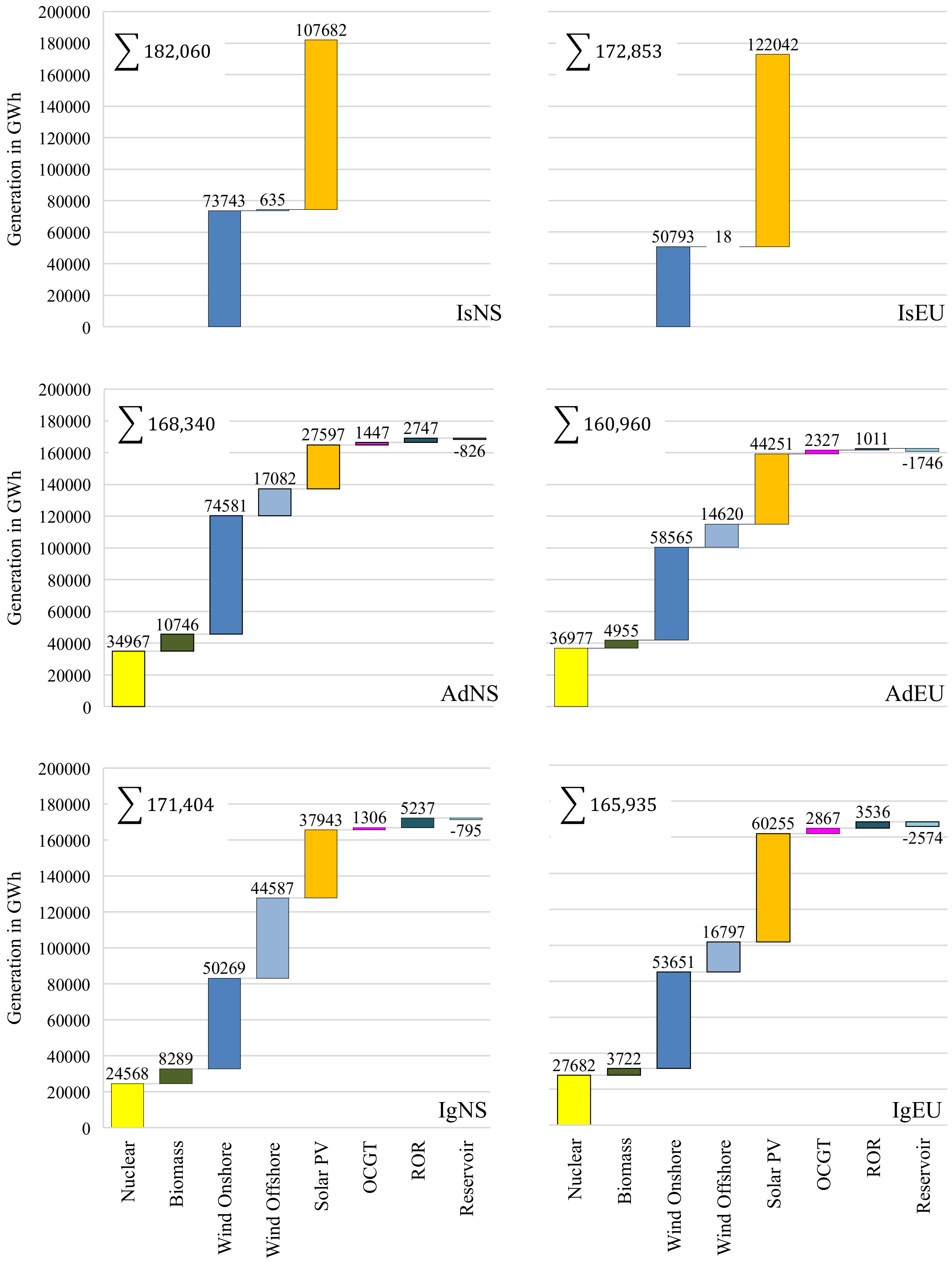

**Figure 4: Scenario-specific electricity generation to power the direct air capture and $CO_2$ storage systems, total electricity generation in the top left of each segment**

The first row of the figure visualises the electricity generation in the Isolated scenarios. As shown by Figure 3, all the electricity consumed by DACCS comes from solar PV and wind (mostly onshore). When $CO_2$-storage is allowed across Europe (IsEU), more PV is used due to the better solar conditions in Southern Europe. Consequently, most direct air systems in IsEU are built on the Iberian Peninsula, which lowers costs through two effects: warmer temperatures reduce the heat pump's electricity demand, and high solar potential minimises the additional capacity needed to meet the electricity demand. Despite building batteries, curtailment of RES is high in the cost-optimal Isolated system. The curtailment share is 14.42 % for onshore wind and 11.25 % for solar PV in IsNS, and 13.21 % (onshore wind) and 11.04 % (solar) in IsEU. This is significantly above the values of the Benchmark scenario, with 12.41 % and 9.99 %, respectively. The model "overbuilds" renewable capacities to keep a high utilisation rate of the direct air capture units, but cannot use the flexibility provided by the "main" power system.

The second row, depicting the Added scenarios, shows a different picture. Compared to the Isolated cases, the electricity mix is more heterogeneous because the model can re-schedule existing assets. This comprises both the use of flexibility, e.g. re-scheduling storages, as well as the use of curtailed energy in the optimal system absent of DACCS. In fact, since little capacity is added beyond the Benchmark configuration (Figure 3), most of the additional energy comes from reduced curtailment or ramping up available generation capacities. For example, the share of onshore wind curtailed in the benchmark without DACCS decreases from 12.41 % to 9.58 % (AdNS) and 9.42 % (AdEU), for solar power from 9.99 % in the benchmark to 7.25 % in AdNS and 7.22 % in AdEU. Nuclear capacity is also ramped up during some hours, as the direct air capture units are now placed close to these potentials in France as shown by Figure 2. A comparison between the North Sea and European storage set-ups, i.e. comparing the left and right graphs in the middle of Figure 4, highlights the role of $CO_2$-storage locations. Restricting storage to the North Sea increases electricity demand by 7,380 GWh (around 5 % in total), as heat pumps require more input, which in turn raises generation needs. This favours offshore wind and hydro, located closer to direct air capture sites. Fewer direct air capture capacities were placed in the south, restricted by additional $CO_2$ transporting costs. By contrast, in the AdEU scenario, the model avoids North Sea storage sites and leverages sites with better renewable potentials and lower direct air capturing electricity requirements. As a result, the generation mix shifts toward more nuclear in France and more solar PV in southern Europe, with less wind, hydro, and overall electricity generation.

The third row shows the Integrated scenarios, which give the model the most flexibility for optimisation. The results are broadly similar to the Added scenarios, suggesting that the Added scenario already captures most of the advantages of the Integrated solution. As in the Added scenarios, the utilisation share of renewable generation capacity is higher in the fully integrated scenarios when compared to the Benchmark scenario, with curtailment shares for wind onshore and offshore and solar PV of 9.92 %, 10.35 %, and 8.64 % for the IgNS; and 10.53 %, 11.12 %, and 7.89 % for the IgEU scenario.

## 4.3 Cost Analysis

As the infrastructure expansion differs significantly between the scenarios, so do the cost components of the carbon capture process, as depicted by Figure 5 below. The diagram illustrates the scenario-specific costs of the DACCS chain, broken down into: (i) direct air capture costs (orange palette) – comprising the heat pump CAPEX, direct air capture unit CAPEX and OPEX, and electricity cost for supplying the demand; (ii) $CO_2$ transport cost including CAPEX and maintenance cost (grey palette); and (iii) $CO_2$ storage cost – including CAPEX, fixed and variable operation and maintenance costs, and electricity cost for supplying the storage (green palette).

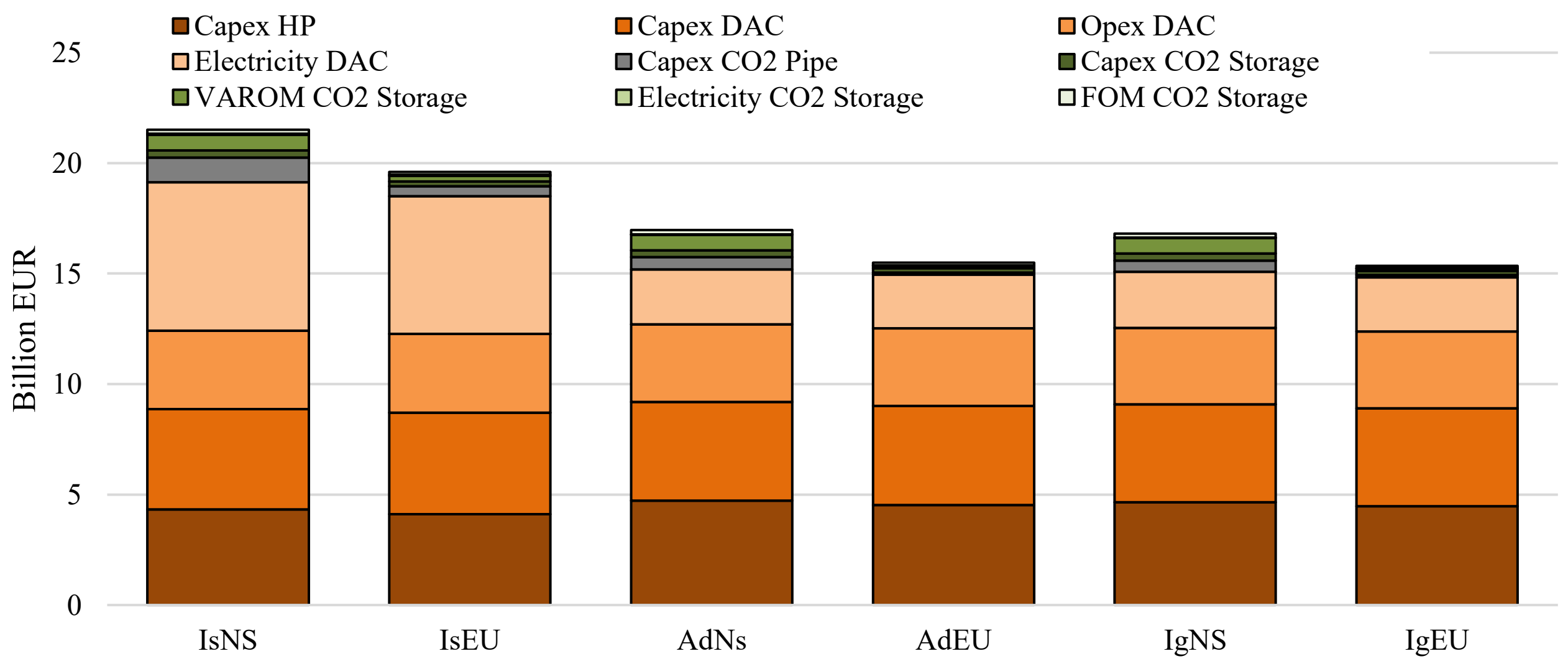

**Figure 5: Scenario-specific DACCS cost distribution**

We can derive three main findings: First, the Isolated scenarios are the most expensive, while the Added scenarios are marginally more costly than their Integrated counterparts. Overall, total DACCS costs are about 30 % higher when direct air capture units are optimised in isolation compared to scenarios that account for the underlying energy system.

Second, the CAPEX and OPEX of the direct air capture process (including the heat pumps), i.e. the three segments at the bottom of each column, remain almost constant across scenarios. This indicates limited optimisation potential on the direct air capture unit's equipment cost. Instead, the differences between the scenarios stem from the expenses for electricity. As the model must build new generation capacity from scratch without making use of existing, unused, or curtailed resources in the Isolated scenarios, the cost of electricity for direct air capture units is more than twice as high (>6 bn EUR p.a.) as in the Added and Integrated scenarios (around 2.5 bn EUR p.a., depending on scenario).

Third, the results show that the "North Sea" set-up is consistently more expensive than the "Europe" set-up. This cost difference – about 10 % – stems from the lower $CO_2$ transport and storage costs in the European case. Moreover, the Isolated scenarios exhibit the highest transportation costs overall, since the regions with favourable renewable potentials are located at greater distances from available storage sites. These key findings can also be expressed as scenario-specific average DACCS and Electricity cost (in EUR/t$CO_2$) as visualised by Figure 6 below.

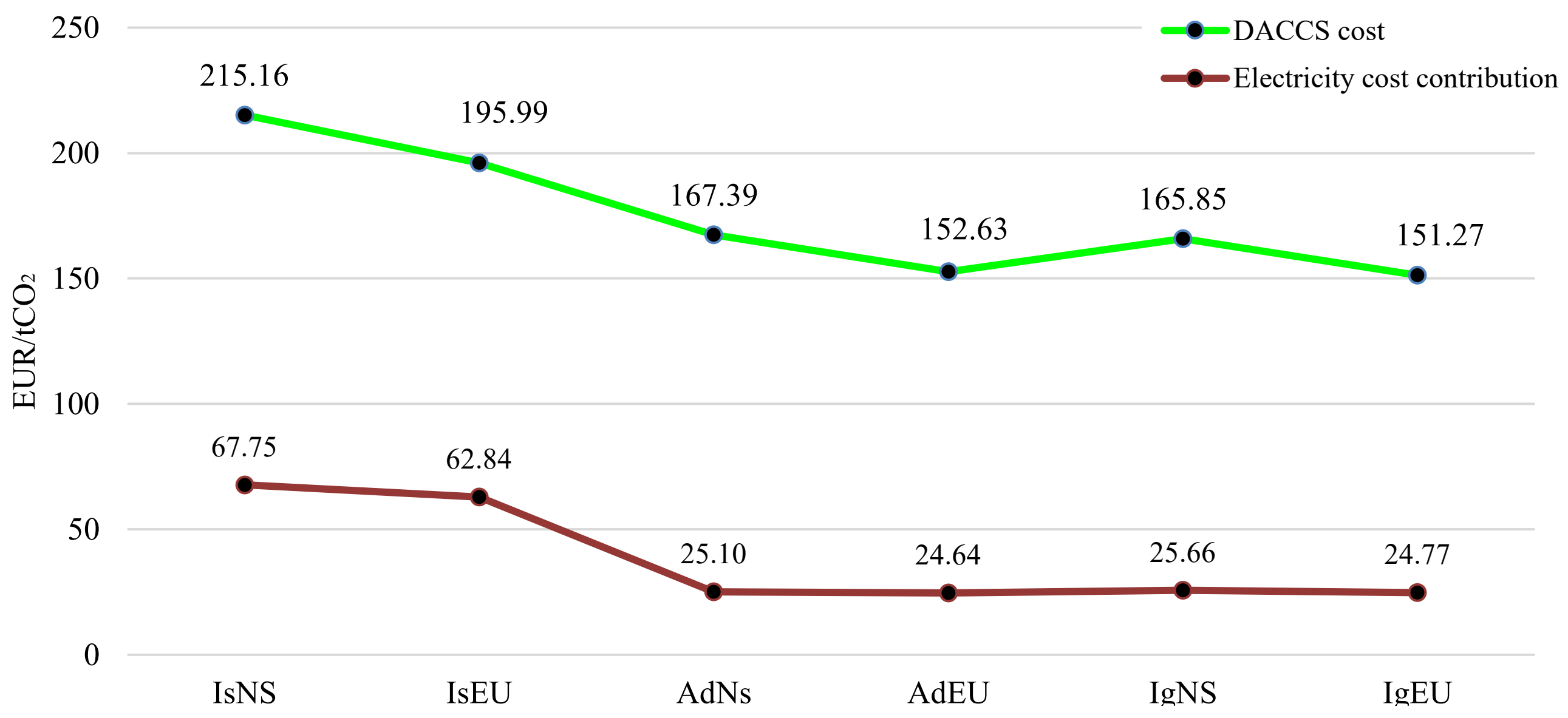

**Figure 6: Scenario-specific DACCS cost and electricity cost per ton $CO_2$ captured**

The average DACCS costs from the Isolated scenarios amount to 215 EUR/t$CO_2$ and 196 EUR/t$CO_2$. In contrast, the Added and Integrated scenarios yield substantially lower costs of 167 and 151 EUR/t$CO_2$. Interestingly, the DACCS cost difference between the Added and Integrated scenarios is very small: lower than 1 %. This indicates a low regret potential from not explicitly integrating DACCS deployment at the presented scale into the planning of a European carbon-free electricity system. Electricity-related costs show a more pronounced difference across scenarios. In the isolated cases, electricity costs reach 68 EUR/ t$CO_2$ and 62 EUR/ t$CO_2$, which is more than twice as high as in the added and integrated scenarios (around 25 EUR/t$CO_2$), where values remain relatively stable across configurations. This suggests that existing system flexibilities are effectively utilized when DACCS is integrated, significantly reducing electricity costs. As a result, the storage location emerges as the dominant driver of cost variation within the systems.

# 5 CONCLUSION

Our research is motivated by the high variation in existing cost estimates for future DACCS deployment, which poses challenges for investors, policymakers, and technology development – particularly given the essential role of negative emissions in meeting climate targets. We identified and explored two key drivers for cost variation: differences in modelling approaches and permitted storage sites. To quantify these impacts on the future DACCS cost, we developed an energy system model that covers the European electricity sector and the DACCS infrastructure, considering the LT direct air capture system as the main capturing technology. We set an exogenous target for $CO_2$ withdrawals (100 Mt $CO_2$ p.a.) and endogenously optimise investment and dispatch decisions. For the direct air capture systems, this comprises optimal placements accounting for local temperature and humidity, renewable availability and generation potentials, grid connections, $CO_2$ storage sites, and transport options.

Taken together, the two differences we explore in this analysis can change capture costs by more than 60 EUR (nearly 30 %) per ton of $CO_2$: in the stand-alone optimisation with storage sites limited to North Sea locations, capture costs amount to 215.16 EUR/t$CO_2$, compared to 151.27 EUR/t$CO_2$ when the DACCS electricity consumption is taken into account when planning the energy system and storage is allowed across Europe. This result thus explains a significant part of the differences in capture costs between studies, as resulting from differences in methodology and storage siting.

The methodological modelling approach has the high individual impact on both the estimated carbon capture cost and the location and optimal energy supply of the direct air capture units. We demonstrate that taking into account system integration, i.e. neither treating direct air capture units as inflexible price takers nor as the sole electricity consumers in a system, reduces capture costs by up to 30 %. Our work thus complements and extends the studies by Fasihi et al. (2019), Young et al. (2023), Terlouw et al. (2024) and Wenzel et al. (2025a), all of which do not take the system perspective. A main driver of this cost reduction is the reduction in new electricity generation capacities required to produce the electricity demand from DACCS. The reduction in CAPEX reduces the electricity cost of the DACSS system when DACCS complements other electricity consumers in an integrated system. Another cost reduction results from the optimal placing of the various DACCS components in the system. This includes the location of storage sites for $CO_2$. When $CO_2$ storage locations are restricted to the North Sea, the cost-optimal solution places direct air capture units in southern Europe and builds pipelines for $CO_2$-transport from the south of Europe to the North Sea. Moreover, by explicitly representing $CO_2$ transport and storage infrastructure expansion, we identify a cost increase of approximately 10 % when storage is restricted to the North Sea. Modelling citing decisions endogenously extends studies relying on uniform transport and storage cost assumptions (e.g. Lux et al. 2023 and Young et al. 2023).

Our findings have several implications for politicians and researchers. First, a cost-optimal DACCS infrastructure involves considering system flexibilities and excess electricity generation when locating direct air capture units. Optimising their placement in isolation leads to sub-optimal results. Second, such optimisation requires either a functioning electricity market with undistorted local price signals or a Europe-wide coordination to optimally place the different components of DACCS systems, lowering costs for all. Third, allowing onshore storage of $CO_2$ within Europe is cheaper and reduces the expansion of $CO_2$ pipelines. This finding is especially interesting for countries that are currently not allowing onshore $CO_2$ storage. However, our results indicate that restricting storage sites to the North Sea, rather than allowing additional onshore options across Europe, does not decisively affect total DACCS costs. This suggests that limiting storage to offshore sites could be a viable option for addressing social acceptance concerns. At the same time, politicians in Europe should thus be aware that North Sea storage would imply the need to build or repurpose pipelines for the $CO_2$ transport across countries, which itself raises new public acceptance challenges.

Our case study is based on a linear cost minimisation model, which comes along with its typical limitations, among which is its sensitivity to input parameter changes, especially on the cost side. Furthermore, modelling the expansion with a mixed-integer approach to better capture grid, pipeline, and direct air capture system expansion would also yield more accurate results. However, this would also render this problem intractable or require major simplifications in other aspects, which in our view would reduce the quality of the empirical results. Additionally, we face other trade-offs between tractability and accuracy. On the electricity side, we assume that battery and hydrogen storage, as well as electrolysers, are built with a fixed energy-to-power ratio. While this is the most widely used approach in the literature on energy systems modelling, it underestimates storage investment flexibility. We also fix the available nuclear and biomass capacity to the current (2025) level within Europe. Other approaches, e.g. allowing for endogenous expansion of these technologies, would influence the results. On the $CO_2$ side, we do not account for HT direct air capture systems, which, when combined with industries or other heat sources, could influence the path to achieving the targeted negative emissions goal in a more cost-effective manner, thereby altering the results. Moreover, our $CO_2$ pipeline costs are computed linearly using distance, without accounting for terrain-specific challenges or construction constraints. Additionally, we assume full connectivity, meaning that any pair of nodes can be linked via a pipeline without spatial or regulatory limitations. For the $CO_2$ storage, we assume available free capacities where the direct air capture's $CO_2$ volumes can be stored. This might not reflect the real available storage capacities, of which volumes are still based on estimates.

From the European perspective, further work could integrate the planned EU PCI projects for $CO_2$-pipelines in the electricity and DACCS expansion planning to investigate the impact on the optimal direct air capture system placement. Also, the consideration of industry storage sites with their capturing potentials and proposed infrastructure projects would be interesting. Furthermore, it would be interesting to analyse regret between scenarios, such as our Added and the Integrated scenarios, when more $CO_2$ must be captured by direct air capture units. What is the price of not including DACCS into the energy system planning for a more ambitious 200, 300, 400 or 500 Mt p.a. negative emission target? Finally, as we focus on the electricity sector, a representation of the heat sector could provide more insights into the system dynamics, especially when also accounting for HT direct air capture systems.

# RESOURCE AVAILABILITY

## Lead Contact

Requests for further information and resources should be directed to and will be fulfilled by the lead contact, Maximilian Bernecker (bernemax@b-tu.de).

## Materials Availability

This study did not generate new unique reagents.

## Data and Code Availability

All original code and data has been deposited at GitHub and is publicly available at https://github.com/bernemax/DACCS-Europe-2050 as of the date of publication.

# AKNOWLEDGEMENTS

Maximilian Bernecker gratefully acknowledges supported by the German Federal Government, the Federal Ministry of Education and Research, and the State of Brandenburg within the framework of the joint project EIZ: Energy Innovation Center (project numbers 85056897 and 03SF0693A) with funds from the Structural Development Act (Strukturstärkungsgesetz) for coal-mining regions."

Felix Müsgens gratefully acknowledges financial support from the Federal Ministry of Education and Research of Germany in the Ariadne project (03SFK5S0).

# DECLATRATION OF INTERESTS

The authors declare no competing interests.

# AUTHOR CONTRIBUTIONS

Maximilian Bernecker: Conceptualization, Methodology, Software, Data curation, Writing – original draft, Visualization, Validation.

Felix Müsgens: Conceptualization, Methodology, Writing – review & editing, Validation, Funding acquisition, Supervision.

# DECLARATION OF GENERATIVE AI AND AI-ASSISTED TECHNOLOGIES

During the preparation of this work the authors used DeepL to improve clarity and writing style of the document. After using this tool/service, the authors reviewed and edited the content as needed and take full responsibility for the content of the published article.

# APPENDIX A

**Table 2: Electricity technology cost data**

| Technology | Lifetime [years] | Interest Rate [%] | Overnight costs | Fixed OM costs | Fuel Costs [EUR/$MWh_{th}$] | Efficiency [%] | Data Source |
|---|---|---|---|---|---|---|---|
| Onshore Wind | 30 | 7.5 | 963 EUR/kW | 9.63 EUR/kW/y | - | - | Danish Energy Agency |
| Offshore Wind | 30 | 9.3 | 1380 EUR/kW | 13.80 EUR/kW/y | - | - | Danish Energy Agency |
| Solar PV | 35 | 6.9 | 370 EUR/kW | 7.40 EUR/kW/y | - | - | Danish Energy Agency |
| Battery Inverter | 10 | 6.0 | 60 EUR/kW | 0.6 EUR/kW/y | - | 96 | Danish Energy Agency |
| Battery Storage | 30 | 6.0 | 75 EUR/kWh | 0.6 EUR/kWh/y | - | - | Danish Energy Agency |
| Electrolyzer | 25 | 8.0 | 350 EUR/kW | 14 EUR/kW/y | - | 70 | Danish Energy Agency |
| $H_2$ Storage Tank | 30 | 8.0 | 21 EUR/kWh | 0.5 EUR/kWh/y | - | - | Danish Energy Agency |
| $H_2$-fired OCGT | 25 | 8.0 | 411 EUR/kW | 8.7 EUR/kW/y | - | 43 | Danish Energy Agency |
| AC Transmission | 40 | 6.0 | 700-2500 EUR/MW/km | - | - | - | NEP 2023 (1)- |
| DC Transmission | 40 | 6.0 | 2500-5500 EUR/MW/km | - | - | - | NEP 2023 (1)- |
| Biomass | 60 | - | - | 80 EUR/kW/y | 4.5 | 47 | DIW (2) |
| Nuclear | 60 | - | - | 100 EUR/kW/y | 3.0 | 33 | DIW (2) |
| Run of River | - | - | - | 35 EUR/kW/y | - | - | DIW (2), IEA |
| Pumping Storage | - | - | - | 45 EUR/kW/y | - | - | DIW (2), IEA |
| Reservoir | - | - | - | 40 EUR/kW/y | - | - | DIW (2), IEA |
| Central air-sourced heat pump | 25 | 8.0 | 906.10 $EUR/kW_{th}$ | 21.17 $EUR/kW_{th}/y$ | | | Danish Energy Agency |

**Table 3: Technical and economical parameters of direct air capture systems**

| Technology | Power Demand [$kWh/tCO_2$] | Heat Demand [$kWh/tCO_2$] | CAPEX [$EUR/(tCO_2/y)$] | Lifetime [year] | Interest rate [%] | FOM [$€/(tCO_2/y)$] | Forced outage rate [%] | References |
|---|---|---|---|---|---|---|---|---|
| Soldi-LT-Direct Air Capture | 204 (105-370) | 1257 (1900-4378) | 315 | 20 | 10% | 29 | 6 | (Wenzel et al., 2025a, 2025b) |

- Temperature and humidity-dependent electricity and heating demand from (Wenzel et al., 2025a)
    - temperature and humidity data for each node and timestep in the system taken from renewable ninja.

- CAPEX, Lifetime, FOM from (Wenzel et al., 2025b)
- Heat is supplied via heat-pump using temperature dependent COP
  - $cop_{hp,t} = \frac{T^{heat}}{(T^{heat} - T^{ambient}_{hp,t})} \cdot \eta$
    - $T^{heat}$ = 383.15 K (110°C)
    - $T^{ambient}_{hp,t}$ time dependent air temperature in K $\forall\ hp \in \Phi^{HP}_{n}$
    - $\eta = 0.5$ (Carnot efficiency)

**Table 4: Technical and economical parameters of $CO_2$ storage**

| Technology | Injection Capacity [Mt/y) | CAPEX [€/t$CO_2$/y] | Fix OM [€/t$CO_2$/y] | Variable OM [€/t$CO_2$] | Power Demand [kWh/t$CO_2$] | References |
|---|---|---|---|---|---|---|
| Onshore Storage | 3 | 1.85 | 1.1 | 0.99 | 13.88 | (Danish Energy Agency, 2024) |
| | 5 | 1.93 | 0.71 | 1.02 | 13.88 | |
| Offshore Storage | 3 | 2.83 | 1.88 | 7.05 | 13.88 | |
| | 5 | 3.18 | 1.19 | 2.51 | 13.88 | |

**Table 5: Technical and economical parameters of $CO_2$ pipeline**

| Technology | Transport Volume [Mt $CO_2$/y] | CAPEX [EUR/(t$CO_2$/km)] | Fix OM [EUR/(t$CO_2$/h)/km] | Lifetime [year] | Interest rate [%] | Reference |
|---|---|---|---|---|---|---|
| Pipeline Onshore | 18 | 0.19 | 20 | 40 | 5 | (Holz et al., 2021) |
| Pipeline Offshore | 18 | 0.38 | 20 | 40 | 5 | |

We estimate the offshore pipelines to be two times more expensive compared to the onshore pipelines, which is in line with estimates from (ZEP, 2011) and (IEA, 2020)

**Table 6: Information storage sites**

| Nr. | Annual CAPEX EUR/t$CO_2$ | FIX O&M costs EUR/t$CO_2$ | Var O&M costs EUR/t$CO_2$ | Energy Demand in kWh/tCO2 | Max Stor in Mt/y | Project, PCI number/ source | |
|---|---|---|---|---|---|---|---|
| 1 | 3.18 | 1.88 | 7.05 | 13.88 | 5 | CO2TransPorts | 13,1 |
| 2 | 3.18 | 1.88 | 7.05 | 13.88 | 22 | Aramis | 13,2 |
| 3 | 3.18 | 1.88 | 7.05 | 13.88 | 10.5 | Bifrost | 13,4 |
| 4 | 3.18 | 1.88 | 7.05 | 13.88 | 20 | EU2NSEA - Smeaheia | 13,8 |
| 4 | 3.18 | 1.88 | 7.05 | 13.88 | 5 | EU2NSEA - Luna | 13,8 |
| 4 | 3.18 | 1.88 | 7.05 | 13.88 | 5 | Northern Lights | 13,13 |
| 5 | 3.18 | 1.88 | 7.05 | 13.88 | 12.5 | Kinno, Iroko, Atlas | |
| 6 | 3.18 | 1.88 | 7.05 | 13.88 | 14 | Trudvang, Albondigas, | |
| 7 | 3.18 | 1.88 | 7.05 | 13.88 | 12 | Poseidon, Havstjerne | https://www.catf.us/de/carbon-capture/storage-project-capacity-europe/ |
| 8 | 3.18 | 1.88 | 7.05 | 13.88 | 20 | Hewett Storage Site | https://www.gov.uk/government/publications/carbon-capture-readiness-co2-storage-site-hewitt-bunter |
| 9 | 3.18 | 1.88 | 7.05 | 13.88 | 10 | Viking CCS Storage Site | https://www.vikingccs.co.uk/ |
| 10 | 3.18 | 1.88 | 7.05 | 13.88 | 6 | Orion Carbon Capture and Storage | https://perenco-ccs.com/the-orion-project/ |
| 11 | 3.18 | 1.88 | 7.05 | 13.88 | 26 | Northern Endurance Partnership | https://northernendurancepartnership.co.uk/2025/06/12/northern-endurance-partnership-welcomes-uk-government-support-for-ccs/ |
| 12 | 3.18 | 1.88 | 7.05 | 13.88 | 25 | Spirit Energy CCUS hub | https://www.spirit-energy.com/our-operations/mnz/ |
| 13 | 1.93 | 1.1 | 0.99 | 13.88 | 6 | PCI | 13,12 |
| 14 | 3.18 | 1.88 | 7.05 | 13.88 | 6.4 | PCI | 13,5 |
| 15 | 1.93 | 1.1 | 0.99 | 13.88 | 8 | Norn | 13,10 |
| 16 | 1.93 | 1.1 | 0.99 | 13.88 | 10 | Norn | 13,10 |
| 17 | 2.42 | 3.07 | 1.34 | 13.88 | 0.62 | PCI | 13,9 |
| 18 | 1.93 | 1.1 | 0.99 | 13.88 | 13 | PCI | 13,11 |
| 19 | 1.93 | 1.1 | 0.99 | 13.88 | 5.3 | CATF announced (https://www.catf.us/resource/unlocking-europes-co2-storage-potential-analysis-optimal-co2-storage-europe/) | |
| 20 | 1.93 | 1.1 | 0.99 | 13.88 | 19 | CATF announced (https://www.catf.us/resource/unlocking-europes-co2-storage-potential-analysis-optimal-co2-storage-europe/) | |

| | | | | | | |
|---|---|---|---|---|---|---|
| 21 | 1.93 | 1.1 | 0.99 | 13.88 | 5 | CATF announced (https://www.catf.us/resource/unlocking-europes-co2-storage-potential-analysis-optimal-co2-storage-europe/) |
| 22 | 1.93 | 1.1 | 0.99 | 13.88 | 8.1 | CATF announced (https://www.catf.us/resource/unlocking-europes-co2-storage-potential-analysis-optimal-co2-storage-europe/) |
| 23 | 1.93 | 1.1 | 0.99 | 13.88 | 3.7 | CATF announced (https://www.catf.us/resource/unlocking-europes-co2-storage-potential-analysis-optimal-co2-storage-europe/) |
| 24 | 1.93 | 1.1 | 0.99 | 13.88 | 24 | CATF announced (https://www.catf.us/resource/unlocking-europes-co2-storage-potential-analysis-optimal-co2-storage-europe/) |
| 25 | 1.93 | 1.1 | 0.99 | 13.88 | 15 | CATF announced (https://www.catf.us/resource/unlocking-europes-co2-storage-potential-analysis-optimal-co2-storage-europe/) |
| 26 | 1.93 | 1.1 | 0.99 | 13.88 | 18 | CATF announced (https://www.catf.us/resource/unlocking-europes-co2-storage-potential-analysis-optimal-co2-storage-europe/) |
| 27 | 1.93 | 1.1 | 0.99 | 13.88 | 8.6 | CATF announced (https://www.catf.us/resource/unlocking-europes-co2-storage-potential-analysis-optimal-co2-storage-europe/) |
| 28 | 1.93 | 1.1 | 0.99 | 13.88 | 20 | CATF announced (https://www.catf.us/resource/unlocking-europes-co2-storage-potential-analysis-optimal-co2-storage-europe/) |
| 29 | 1.93 | 1.1 | 0.99 | 13.88 | 16 | CATF announced (https://www.catf.us/resource/unlocking-europes-co2-storage-potential-analysis-optimal-co2-storage-europe/) |
| 30 | 2.42 | 3.07 | 1.34 | 13.88 | 1.5 | CATF announced (https://www.catf.us/resource/unlocking-europes-co2-storage-potential-analysis-optimal-co2-storage-europe/) |
| 31 | 1.93 | 1.1 | 0.99 | 13.88 | 8.9 | CATF announced (https://www.catf.us/resource/unlocking-europes-co2-storage-potential-analysis-optimal-co2-storage-europe/) |
| 32 | 1.93 | 1.1 | 0.99 | 13.88 | 8 | CATF announced (https://www.catf.us/resource/unlocking-europes-co2-storage-potential-analysis-optimal-co2-storage-europe/) |
| 33 | 1.93 | 1.1 | 0.99 | 13.88 | 17 | CATF announced (https://www.catf.us/resource/unlocking-europes-co2-storage-potential-analysis-optimal-co2-storage-europe/) |

# APPENDIX B

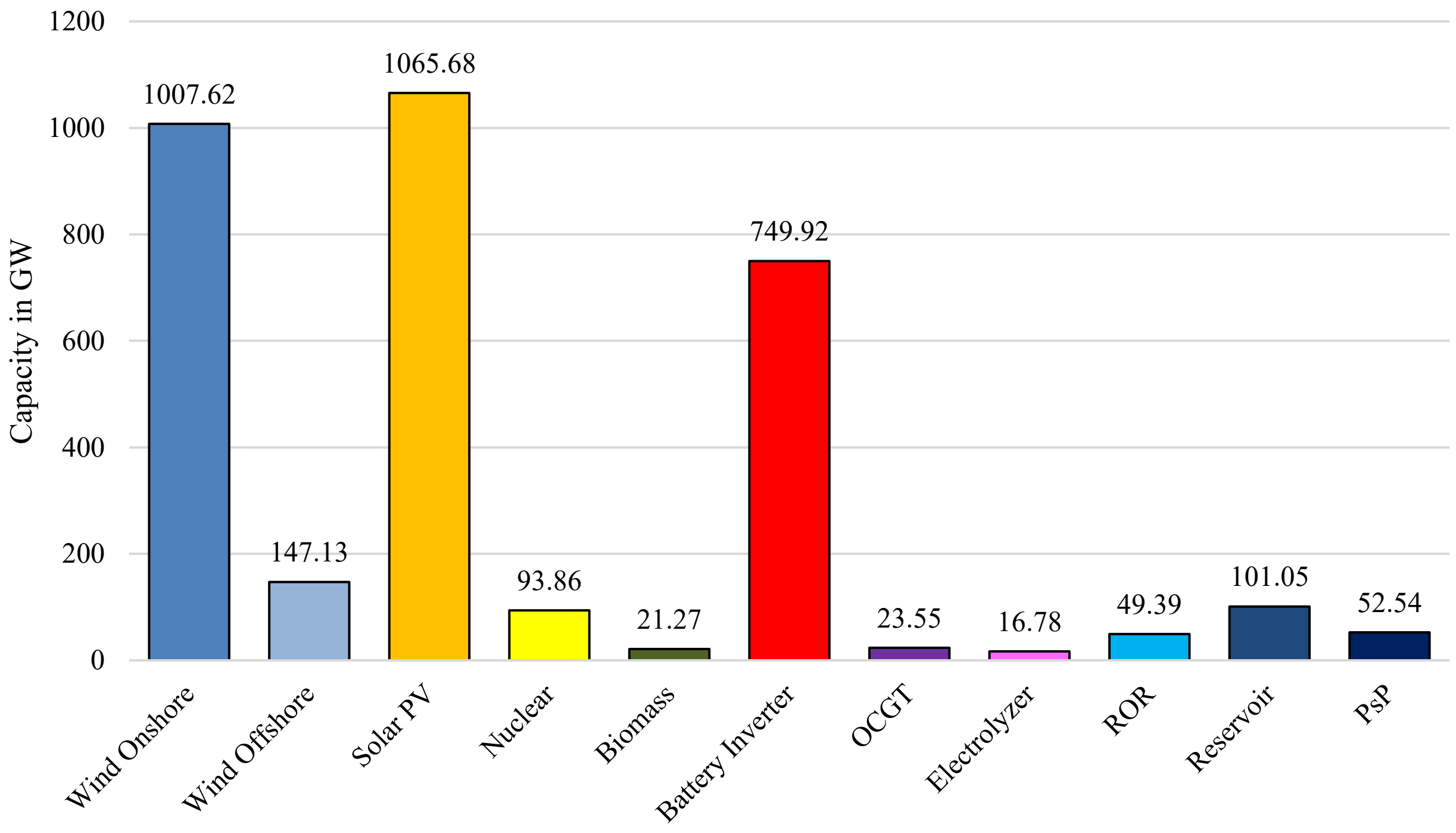

**Figure 7: Benchmark system generation capacity mix**

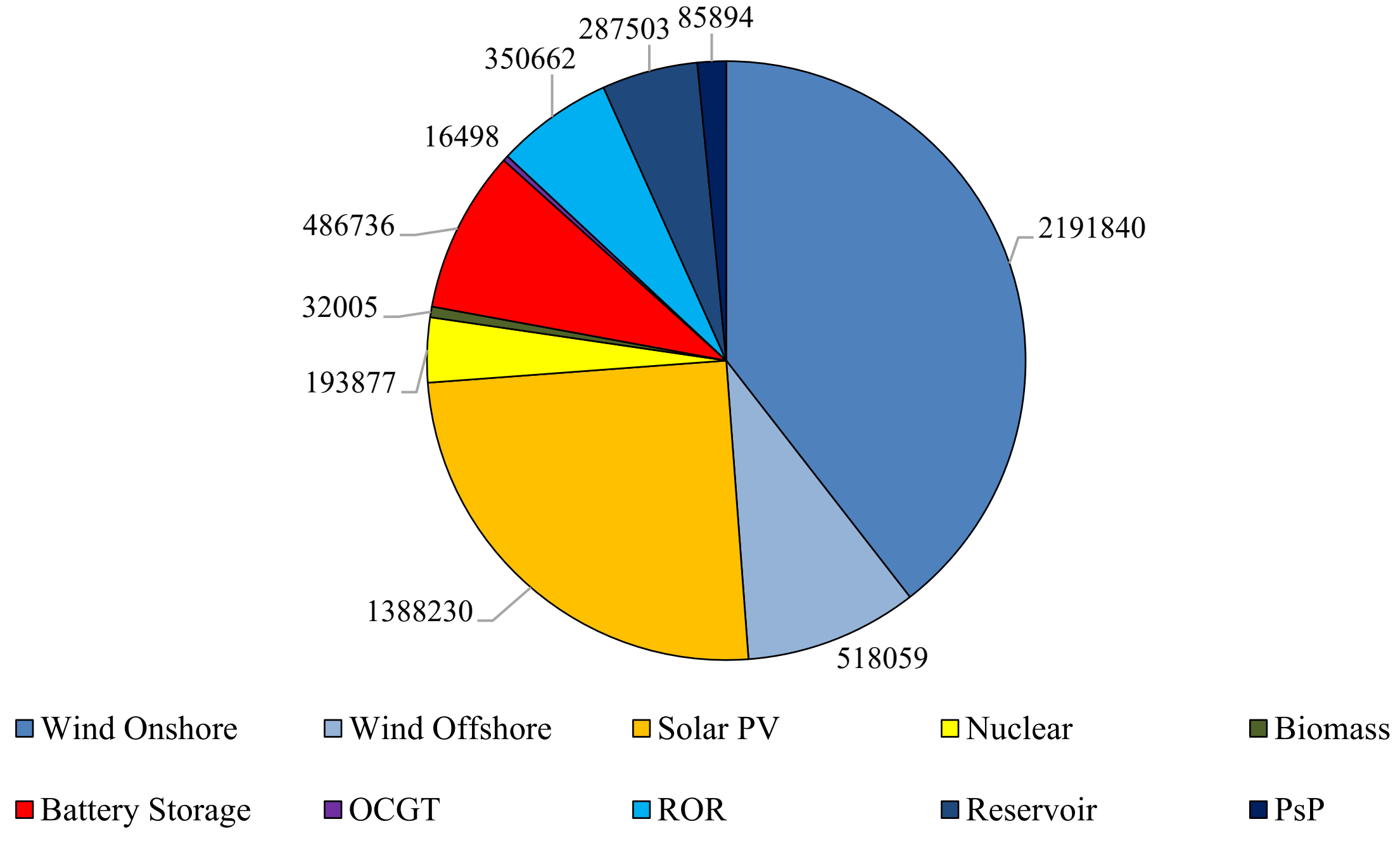

**Figure 8: Benchmark system generation mix in MWh**

# APPENDIX C

**Detailed description of Figure 2:**

Comparing North Sea storage (left column) with European storage options (right column) reveals that the option to store in locations besides the North Sea is used by the model: $CO_2$-sinks are more evenly distributed across Europe. In fact, hardly any North Sea storage is used when other options are allowed. Furthermore, NS-scenarios consistently "force" the model to transport $CO_2$ over relatively large distances, as optimal storage location (North Sea) has limited low-cost electricity supply for direct air capture systems.

By comparing the individual figure segments, it becomes clear that there is a fundamental impact of the two scenario dimensions, modelling approach, and the storage location on the system configuration. We can see that the modelling approach differentiates the Isolated scenarios from the Added and Integrated scenarios and the storage location in the offshore from the European scenarios. Starting in the top left with the IsNS scenario, most of the direct air capture capacity is installed in Spain, followed by the UK and some minor capacities along the $CO_2$ pipeline network totalling up to 13,950 $tCO_2/h$. The model leverages Spain's and the UK's renewable potential to meet the direct air capture unit's energy demands while establishing transport routes to the North Sea. One major vector leads from Spain, over France and Belgium towards the Netherlands where the connection to the closest offshore storages with a capacity of 27 Mt $CO_2$/year is made (Nr. 1, 2). As those capacities are limited, a second transport vector leads over Germany and Denmark, where another offshore storage is connected to the network with a yearly capacity of 10.5 Mt $CO_2$/year (Nr. 3). From there the pipeline branches off to connect Sweden and Norway with Denmark, transporting the $CO_2$ to the large storage locations close to the west coast of Norway with an additional storage capacity of 29.2 Mt $CO_2$/year in total (Nr. 4, 5, 7). In parallel, a smaller pipeline network is built in the southern UK, transporting the captured $CO_2$ to two domestic storages – a larger site near the west coast (Nr. 12) and a smaller one near the east coast (Nr. 10) – with a combined capacity of 33.3 Mt $CO_2$/year. We can see a different picture if the model is allowed also to utilise the onshore storage locations in the IsEU scenario. Here the direct air capture capacity is completely located in the Iberian Peninsula, primarily again in Spain and some capacity also in Portugal. Compared to the IsNS scenario, the total capacity is slightly higher with 14,060 $tCO_2/h$ in total. The model now completely avoids

North Sea storage sites, as they are more expensive to connect and operate. Thus, the model locates the direct air capture units to the best solar PV potentials across Europe and minimizes the distance of the pipeline network to the closest storage sites. This leads to three major $CO_2$ transport routes. One from the Iberian Peninsula towards the north to a storage site in France with a capacity of 15 Mt $CO_2$/year (Nr. 22) and another connection towards the south of France to connect a large storage site in Switzerland with 38 Mt $CO_2$/year (Nr. 14). The third transport route connects Portugal's direct air capture potential and its domestic $CO_2$ storage location with an expanded capacity of 20 Mt $CO_2$/year to the pipeline network in southern France.

In the second row, the expansion results of the Added scenarios are depicted. Comparing across the storage-location dimension reveals two other fundamentally different system configurations. In the AdNS scenario, the model can exploit the entire electricity system (based on the Benchmark set-up) and its inherent flexibilities from diverse supply options and grid connections to power the direct air capture units –at a total capacity of 13,611 t$CO_2$/h. Unlike the IsNS scenario, AdNS develops a transport network with two main corridors: one originating from south-eastern Europe and one from north-eastern Europe, both oriented towards the North Sea storage sites. The network connects direct air capture capacities in Italy, Albania, and Greece to the Dutch offshore storage sites with a combined capturing capacity of 27 Mt $CO_2$/year (Nr. 1, 2). Another route connects direct air capture capacities in North Macedonia, Serbia, and Hungary also to the Dutch and Danish storage sites. Additional capacities from direct air capture systems in France are connected along the Italian transport vector. The North-east transport corridor connects direct air capture capacities in Norway and Finland to the Norwegian storage sites, which add up to a capturing capacity of 37.2 Mt $CO_2$/year in total (Nr. 4, 5, 7). Additional direct air capture systems are located on this network in Denmark and the north of Germany, mainly transporting the $CO_2$ to the Norwegian and Danish domestic storage sites with a capacity of 10.5 Mt $CO_2$/year (Nr. 3). As in the IsNS scenario, a parallel network is built in the UK, connecting the direct air capacities to two storage sites with 15.2 Mt and 10 Mt $CO_2$/year (Nr. 10, 12). However, in contrast to IsNS, less-direct air capacity is built in AdNS, with installations shifting to the north and west of the UK. Also, when onshore storage sites are allowed, the AdEU scenario leads to a similar finding as in the IsEU: no North Sea storage sites are developed. However, the resulting $CO_2$ infrastructure is markedly different, becoming far more decentralised and fragmented. Instead of a single interconnected network in the south-west, several regional clusters emerge in which capture, transport, and storage take place independently with a total direct air capture capacity of 13,779 $tCO_2$/h. In the north-east, direct air capture units in Latvia and Lithuania are developed close to the storage sites at 2.5 Mt $CO_2$/year and 2.2 Mt $CO_2$/year respectively (Nr. 30, 29). Furthermore, direct air capture capacities in Poland are linked to a domestic storage site with a capacity of 8.8 Mt $CO_2$/year (Nr. 32). A south-eastern cluster connects direct air capture units in Hungary, Albania, North Macedonia, and Greece to storage sites in Hungary (4.8 Mt $CO_2$/year, Nr. 26), Croatia (3.2 Mt $CO_2$/year, Nr. 25), Bulgaria (8.9 Mt $CO_2$/year, Nr. 13), and Greece (13 Mt $CO_2$/year, Nr. 24). In Italy, southern direct air capture units are located close to the storage site with a capacity of 6.7 Mt $CO_2$/year (Nr. 27). Additionally, direct air capture systems in the mid of Italy are connected to a storage site in Switzerland with a capacity of 11.0 Mt $CO_2$/year. In the north-central region, direct air capture units in Denmark are connected to both domestic sites and to a German site, together providing 22.9 Mt $CO_2$/year storage capacity (Nr. 15, 23). Other capturing hubs are also situated directly adjacent to storage sites including two in France with 5.7 Mt and 10.6 Mt $CO_2$/year (Nr. 22, 23) and one connected to a smaller storage site of 1.1 Mt $CO_2$/year in Ireland (Nr. 37).

Comparing the Integrated scenarios depicted in the third row of Figure 2 with the previous Added scenarios shows that overall, the system configurations are almost the same, with only minor differences in installed storage and direct air capture systems capacities. The reason is that in addition to the existing fixed generation capacity from nuclear, biomass, and hydro, the model now optimises the entire energy system from scratch, explicitly accounting for the additional electricity demand of direct air capture units. This results in a geographic shift of renewable generation capacities – in particular solar PV – from northern to southern regions to cover the additional demand. In the IgNS scenario the pipeline network in the UK differs slightly, with more direct air capture and storage capacity connected. Also, more direct air capture capacity is built in the south-east of Europe, but less in the north-east. In the IgEU scenario no storage capacity is developed in Ireland. Moreover, direct air capture and storage capacity shifts to the south and south-east, accompanied with additional storage and pipeline connections in the south-east. Overall, the system configuration varies strongly across the scenario dimensions – whether direct air capture is optimised in isolation or within the broader energy system, and whether other storage sites are permitted.

# APPENDIX D

The following figures are visualising the carbon absorption process of the direct air capturing unit of the respective scenarios during the course of the year.

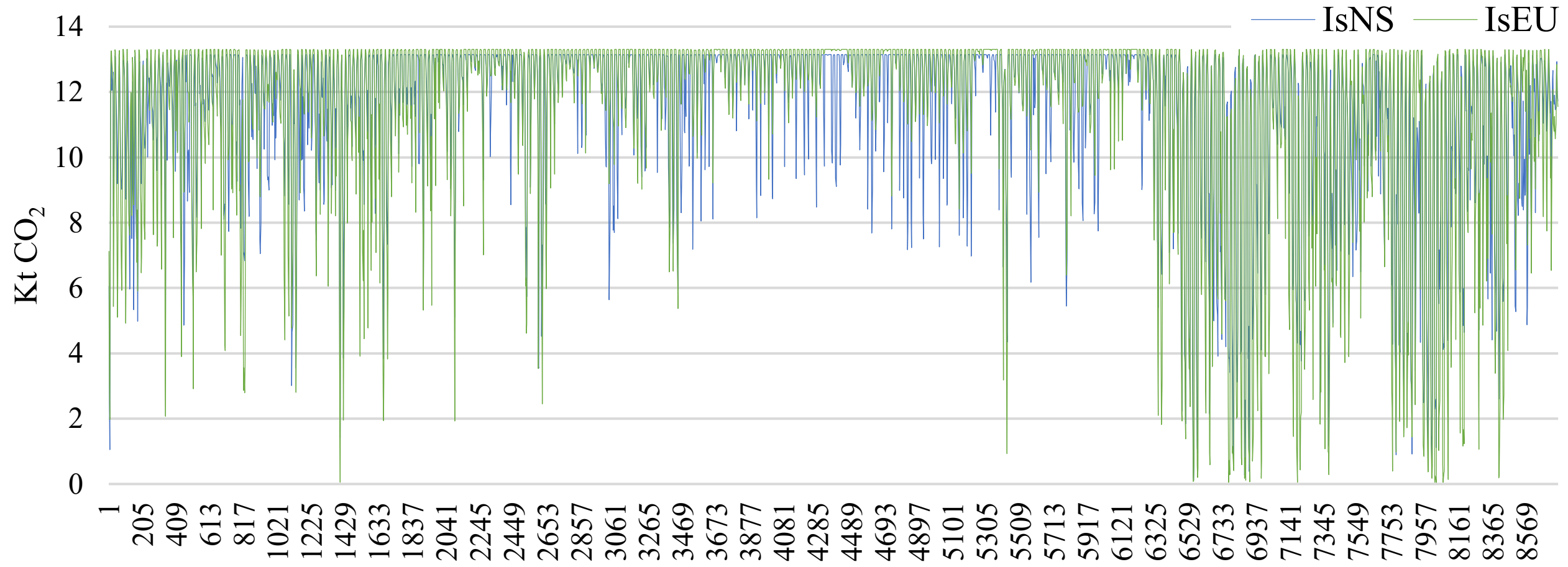

**Figure 9: Capturing profile of direct air capturing units in the IsNS and IsEU scenario**

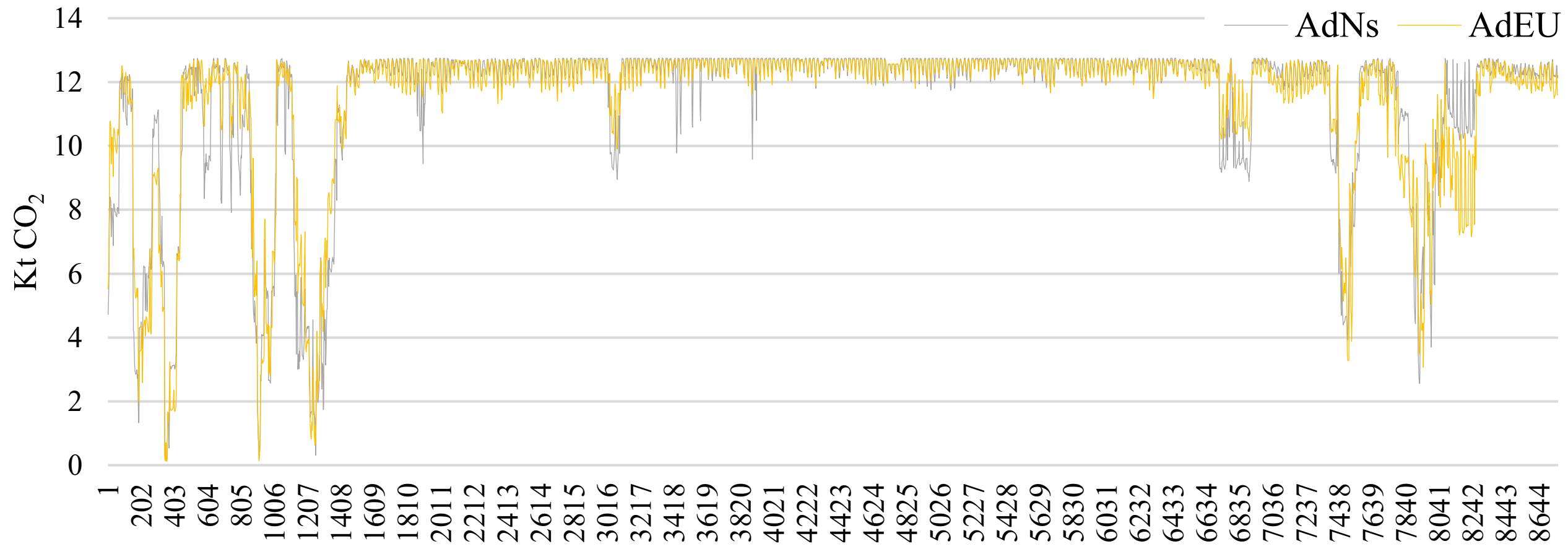

**Figure 10: Capturing profile of direct air capturing units in the AdNS and AdEU scenario**

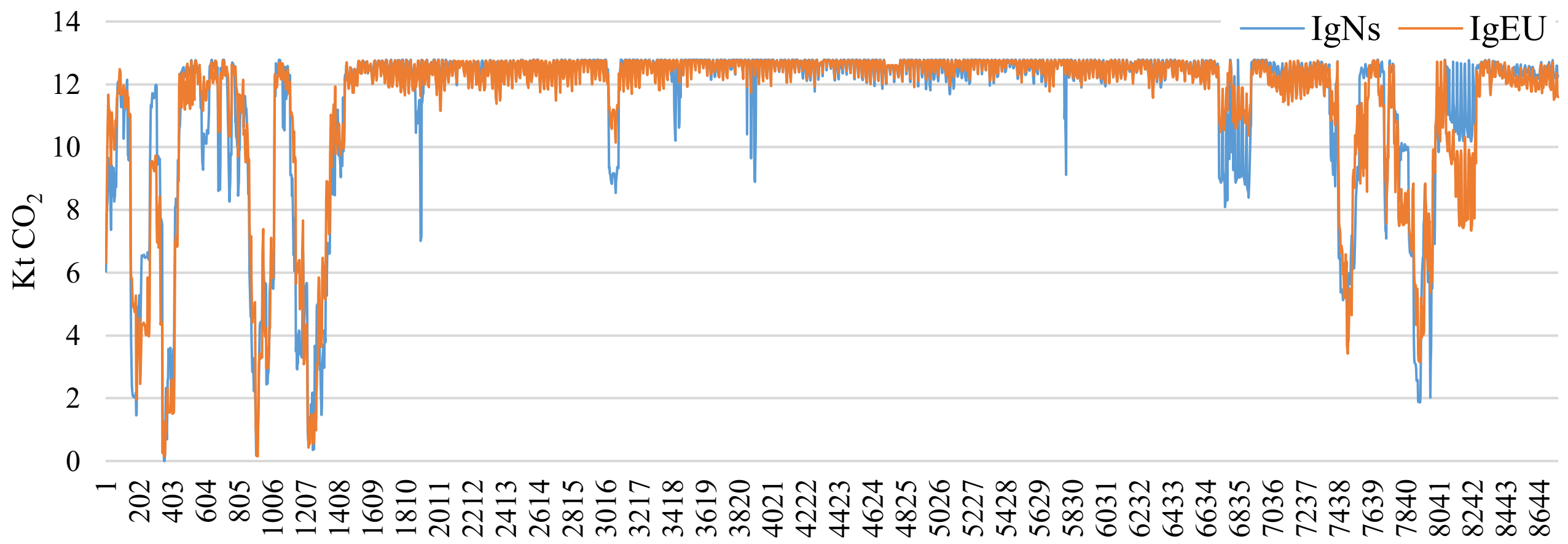

**Figure 11: Capturing profile of direct air capturing units in the IgNS and IgEU scenario**